\documentclass[aps,prc,showpacs,superscriptaddress,nofootinbib,onecolumn]{revtex4-1} 

\usepackage{textcomp} 
\usepackage{amssymb}  
\usepackage{bm}       
\usepackage{booktabs} 
\usepackage{dcolumn}  
\usepackage{graphicx} 
\usepackage[printonlyused]{acronym} 

\usepackage{verbatim} 

\usepackage{graphicx}
\usepackage{capt-of}
\usepackage{amsmath}

\usepackage{relsize}

\usepackage{xcolor}

\usepackage{multirow}
\usepackage{tabularx}

\usepackage{mathrsfs,amssymb} 

\newcommand{\beq}{\begin{equation}}
\newcommand{\eeq}{\end{equation}}
\newcommand{\bes}{\begin{split}}
\newcommand{\ees}{\begin{split}}
\newcommand{\bea}{\begin{eqnarray}}
\newcommand{\eea}{\end{eqnarray}} 
\newcommand{\nn}{\nonumber \\ }
\newcommand{\bfi}{\begin{figure}}
\newcommand{\efi}{\end{figure}}

\renewcommand{\mathbf}[1]{\ensuremath{\boldsymbol{#1}}}

\newcommand{\mathds}[1]{\it \text{\usefont{U}{dsrom}{m}{n}#1}}

\begin{document}

\title{$^6$He Nucleus in Halo Effective Field Theory}

\author{C.~Ji}
\email{jichen@triumf.ca}
\affiliation{Institute of Nuclear and Particle Physics and Department of Physics and Astronomy, Ohio University, Athens, Ohio 45701 USA}
\affiliation{TRIUMF, 4004 Wesbrook Mall, Vancouver, BC V6T 2A3, Canada\footnote{Permanent Address}}

\author{Ch.~Elster}
\email{elster@ohio.edu}
\affiliation{Institute of Nuclear and Particle Physics and Department of Physics and Astronomy, Ohio University, Athens, Ohio 45701 USA}

\author{D.~R.~Phillips}
\email{phillid1@ohio.edu}
\affiliation{Institute of Nuclear and Particle Physics and Department of Physics and Astronomy, Ohio University, Athens, Ohio 45701 USA}

\date{\today}

\begin{abstract}
\begin{description}
\item[Background] In recent years properties of light rare isotopes have been measured with 
high accuracy. At the same time, the theoretical description of
light nuclei has made enormous progress, and properties of, e.g., the 
helium isotopes can now be calculated {\it ab initio}.  These advances make those rare isotopes an ideal testing ground for effective field theories (EFTs) built upon cluster degrees of freedom.

\item[Purpose] Systems with widely separated intrinsic scales are well suited to an EFT treatment.
The Borromean halo nucleus $^6$He exhibits
such a separation of scales. In this work an EFT in which the degrees of freedom are the valence neutrons ($n$) and an inert $^4$He-core
($\alpha$) is employed. The properties of ${}^6$He can then be 
calculated using the momentum-space Faddeev equations for the $\alpha nn$ bound state to obtain information
on ${}^6$He at leading order (LO) within the EFT.

\item[Results] The $nn$ virtual state and the $^2$P$_{3/2}$ resonance in $^5$He give
the two-body amplitudes which are input to our LO
three-body Halo EFT calculation. We find
that without a genuine three-body interaction the two-neutron separation energy $S_{2n}$ of ${}^6$He is strongly
cutoff dependent. We introduce a $nn \alpha$ ``three-body'' operator which
renormalizes the system, adjusting its coefficient to reproduce the $S_{2n}$ of $^6$He. The Faddeev components are then cutoff independent for
cutoffs of the order of, and above, the breakdown scale of the Halo EFT.

\item[Conclusions] As in the case of a three-body system where only resonant
s-wave interactions are present, one three-body input is required for the
renormalization of the EFT equations that describe $^6$He at LO. However, in contrast to the
s-wave-only case, the running of the LO $nn\alpha$ counterterm does not exhibit
discrete scale invariance, due to the presence of the p-wave $n\alpha$ interaction.

\end{description}
\end{abstract}


\maketitle 


\section{Introduction}
\label{sec:intro}

The ${}^6$He nucleus is a prominent example of a ``halo
nucleus"~\cite{Tanihata:1995yv,Jensen:2004,Zhukov:1993aw}. Its two-neutron separation
energy, $S_{2n}= 0.975$ MeV, which is much less than the excitation energy of ${}^4$He, $E^*_{\alpha} \approx 20$ MeV. The last
two neutrons in ${}^6$He thus exist in states whose probability distribution extends
well beyond that of the ${}^4$He core. This encourages a treatment of ${}^6$He as an
effective three-body problem, with ${}^4$He and the two valence neutrons as degrees of 
freedom. In these terms ${}^6$He is a Borromean system, since none of its two-body
subsystems are bound, and the existence of the ${}^6$He
bound state is a genuine three-body phenomenon.
Other neutron-rich nuclei including ${}^{11}$Li,
${}^{22}$C~\cite{Mazumdar:2000dg}, and, perhaps, ${}^{62}$Ca~\cite{Hagen:2013jqa}, can
 also be viewed as Borromean systems.

However, ${}^6$He is special,  since  it is today
accessible to {\it ab initio} methods which compute its structure directly from a 
Hamiltonian which contains state-of-the-art two-nucleon and three-nucleon 
interactions~\cite{Pieper:2004qh,Quaglioni:2013kma,Bacca:2012up}. These calculations confront
experimental data on $S_{2n}$~\cite{Brodeur:2012zz} and the charge~\cite{Wang:2004ze} and matter radii~\cite{Tanihata:1992wf,Alkhazov:1997zz,Kiselev:2005} of ${}^6$He. 
Thus ${}^6$He provides an ideal testbed to study the extent to which an effective
cluster description of the halo dynamics captures essential properties of this nucleus, and
when  {\it ab initio} methods are absolutely necessary.

Descriptions of ${}^6$He in a three-body ansatz have traditionally been implemented in
models, with the $n \alpha$ and $nn$ potentials determined by fitting the
observed properties of the two-body subsystems. In particular, the low-energy ${}^1$S$_0$ $nn$ phase shift and the ${}^2$S$_{1/2}$, ${}^2$P$_{1/2}$ and ${}^2$P$_{3/2}$ $n \alpha$ 
phase shifts were taken into consideration. In the early 1970s, much work was devoted to
this topic,  
with Ghovanlou and Lehman studying in detail which features of these phase shifts have
an impact on the ${}^6$He binding energy~\cite{Ghovanlou:1974zza,Lehman:1982zz}. They found that a model which only
includes the $nn$ resonance in the ${}^1$S$_0$ channel and the ``${}^5$He" resonance in
the ${}^2$P$_{3/2}$ $n \alpha$ channel leads to overbinding of ${}^6$He. The binding
energy could be reduced by including other channels. Three-body cluster models
of ${}^6$He as a $nn\alpha$ system which included more sophisticated input for the $n
\alpha$ and $nn$ potentials were constructed in Refs.~\cite{Funada:1994,Varga:1994};
two-neutron separation energies ranging from $0.68$ to $0.99$ MeV were found. 

Cluster descriptions of halo systems are now enjoying a renaissance, thanks to the application of effective field theory (EFT) methods to these systems. EFT 
provides a systematic expansion in a ratio of low- to high-momentum scales. Halo nuclei enjoy
a separation of these scales, since there is a low-momentum scale,
$M_{lo}$, associated with the binding of the valence neutrons, while the
high-momentum scale, $M_{hi}$, is set by the excitation energy of the nuclear core.

Consequently, in the case of systems where all three participating particles interact in s-waves, two-neutron halo nuclei share universal features with the three-nucleon system~\cite{Bedaque:1999ve}, the ${}^4$He trimer~\cite{Bedaque:1998kg}, and cold atomic systems where three atoms interact near a Feshbach resonance. For a review of this connection see Ref.~\cite{Braaten:2004rn}.  Particularly exciting is the possibility that the Efimov physics~\cite{Efimov70}, having been seen experimentally in recombination rates in cold atomic gases~\cite{Gross:2009,Pollack:2009,Kraemer:2006,Zaccanti:2009}, could also exhibit its existence in halo nuclei~\cite{Hammer:2010kp}. A variety of s-wave 2$n$ halos (e.g. $^{12}$Be and $^{20}$C) were investigated at leading~\cite{Canham:2008jd} and next-to-leading~\cite{Canham:2009xg} order in the $M_{lo}/M_{hi}$ expansion by Canham and Hammer. Recently Hagen {\it et al.} proposed that ${}^{62}$Ca could 
also be an s-wave halo that displayed Efimovian features~\cite{Hagen:2013jqa}. 
The existence and universal features of s-wave 2$n$ halos have also been studied by Yamashita {\it et al.} in a renormalized zero-range model~\cite{Yamashita:2007ej,Frederico:2012xh}. Electromagnetic properties of neutron halos were analyzed in the EFT framework in Refs.~\cite{Rupakprl,Fernando:2011ts,Hammer:2011ye,Rupak:2012,Acharya:2013,Hagen:2013B,Zhang:2013kja}.

In contrast to three-body systems including only s-wave interactions, two of the three pairwise interactions in $^6$He are dominated by p-wave interactions. The $n \alpha$ interaction has a low-energy resonance in the ${}^2$P$_{3/2}$ partial wave, as well as an enhanced phase shift in the ${}^2$P$_{1/2}$ where the resonance is much broader. The first EFT treatment of $n \alpha$ scattering was carried out by Bertulani {\it et al.}~\cite{Bertulani:2002sz}, who treated both the p-wave scattering volume $a_1$, and the p-wave effective ``range", $r_1$, as unnaturally enhanced---i.e., they assumed two fine tunings ($a_1 \sim 1/M_{lo}^3$, $r_1 \sim M_{lo}$). In contrast, Bedaque {\it et al.}~\cite{Bedaque:2003wa} showed that the ${}^2$P$_{3/2}$ $n \alpha$ resonance could be well described by the power counting of Ref.~\cite{Pascalutsa:2002pi}, where the resonance's width is only re-summed in its immediate vicinity. 
Therefore they assigned only the scaling necessary to have a low-energy resonance:
$a_1 \sim 1/(M_{lo}^2 M_{hi})$, $r_1 \sim M_{hi}$, thereby
requiring only one fine tuning.
It is this counting we will use in our present study. 
We observe that the $n \alpha$ ${}^2P_{3/2}$ scattering parameters $a_1=-62.951$
fm$^3$ and $r_1=-0.8819$ fm$^{-1}$~\cite{Arndt:1973} are consistent with the low- and
high-momentum scales $M_{lo}=\sqrt{m_N S_{2n}} \approx 30$ MeV and $M_{hi} \approx
\sqrt{m_N E^*_{\alpha}}=140$ MeV in ${}^6$He ($m_N$ denotes the nucleon mass).\footnote{A more recent analysis of $n-\alpha$ data gives $a_1=-65.7$ fm$^3$, $r_1=-0.84$ fm$^{-1}$~\cite{GHale}. We have checked that using these values instead of those of Ref.~\cite{Arndt:1973} produces only very small differences in our results. Any such differences are certainly smaller than the intrinsic uncertainty in our leading-order calculation.}

In contrast, the recent paper of Rotureau and van Kolck~\cite{Rotureau:2012yu} adopted
the power counting of Ref.~\cite{Bertulani:2002sz}, and then applied the Gamow shell model 
to solve ${}^6$He as a three-body problem. In our conclusion we will compare our results with those of Ref.~\cite{Rotureau:2012yu}.

Another recent study of the three-body problem with resonant pairwise p-wave interactions, which employed the power counting 
of Ref.~\cite{Bertulani:2002sz},
was carried out by Braaten {\it et al.}~\cite{Braaten:2011vf}. These authors attempted to find a scale-free situation in the two-body problem, and examine the corresponding behavior in the three-body problem. In order to do so they took a p-wave ``unitary limit" $|a_1|\rightarrow\infty$ and $r_1\rightarrow 0$. However, as pointed out by Nishida~\cite{Nishida:2011np} (see also Ref.~\cite{Jona-Lasinio:2008}) this p-wave unitary limit is not physical: It yields a two-body spectrum in which one low-energy state has negative norm. Thus the discrete scale invariance discovered by Braaten {\it et al.} in the corresponding three-body problem cannot be realized in nature. This provides strong motivation for us to employ the ``narrow resonance" power counting $a_1 \sim 1/(M_{lo}^2 M_{hi})$, $r_1 \sim M_{hi}$ in our work.

In Sec.~\ref{sec:2body-he6}, we discuss the properties of the $nn$ and $n \alpha$
interactions employed in our work together with their low-energy expansions based on
this power counting. We explain the EFT renormalization procedures which allow us to
start from a two-body interaction and obtain the pertinent t-matrices. These t-matrices
are then inserted into three-body Faddeev equations, for which we solve the homogeneous
version in order to determine the ${}^6$He ground state energy. In
Sec.~\ref{sec:Jacobi-he6} we discuss the spin and angular-momentum coupling of the
three particles which leads to a $0^+$ state of ${}^6$He.
In Sec.~\ref{sec:3body-he6} we present the calculation of $^6$He as a three-body system, building the general Faddeev equations for one spinless particle and two identical fermions, and then projecting them onto the angular-momentum channels which are relevant for the ground state of $^6$He. We find that the ground-state energy is not determined by two-body input alone. 
Instead, it depends strongly on the cutoff in the three-body equations. Thus a $nn \alpha$ contact interaction is mandatory at LO in this EFT. Finally, in Sec.~\ref{sec:concl-he6}, we summarize and discuss our results.


\section{Halo EFT in the two-body sector}
\label{sec:2body-he6}

In this section, we discuss the EFT expansion that we use for the $nn$ and $n \alpha$ interactions,. We develop the LO two-body t-matrices, which encodes the two-body input for our three-body calculation. We discuss the regularization and renormalization procedures in both cases.

\subsection{Halo EFT and effective-range expansions of $nn$ and $n\alpha$ t-matrices}
\label{sec:2body-he6-lo}

Here we write the Lagrangian pertaining to ${}^6$He in terms of our effective $nn\alpha$ degrees of freedom. The $nn$ part of the theory was developed as the pionless EFT in Refs.~\cite{Birse:1998dk,Kaplan:1998tg,vanKolck:1998bw}. Successes of the pionless EFT in the nucleon-nucleon sector are summarized in the reviews ~\cite{Beane:2000fx,Bedaque:2002mn}. The  $n \alpha$ part of the theory was first written down in Ref.~\cite{Bertulani:2002sz} (cf. Ref.~\cite{Hammer:2011ye}). The formulation used here follows that of Ref.~\cite{Zhang:2013kja}. 

We write the Lagrangian $\mathscr{L}$ as a sum of one-body, two-body, and three-body terms.
\begin{equation}
\mathscr{L} =  \mathscr{L}_1 + \mathscr{L}_2 + \mathscr{L}_3~.
\end{equation}
The one-body Lagrangian $\mathscr{L}_1$ is
\begin{equation}
\mathscr{L}_1 
= n^{\dagger} \left( i\partial_0 + \frac{\nabla^2}{2m_n} \right) n 
+ \alpha^{\dagger} \left( i\partial_0 + \frac{\nabla^2}{2m_\alpha} \right) \alpha~, 
\label{eq:L1}
\end{equation}
where $\alpha$ is the spinless field of ${}^4$He with mass $m_\alpha$, and $n^{\dagger}$ is the two-component spinor field of the valence neutron $n^{\dagger} = \left(\begin{smallmatrix} n_\uparrow \\ n_\downarrow \end{smallmatrix}\right)$ with mass $m_n$.
The two-body Lagrangian $\mathscr{L}_2$ include the $nn$ s-wave and $n\alpha$ p-wave interactions:
\begin{align}
\mathscr{L}_2 
=& \eta_0\, s^{\dagger} \left( i\partial_0 + \frac{\nabla^2}{4m_n} -\Delta_0 \right) s 
+ \eta_1\, \pi^{\dagger} \left( i\partial_0 + \frac{\nabla^2}{2(m_n+m_\alpha)} -\Delta_1 \right) \pi
\nn
&+
g_0
\left[ s^{\dagger} T_0^{\sigma \delta} n_{\sigma} n_{\delta} + {\rm h.c.} \right]
+  
g_1
\left[ \pi^{\dagger a} T_{a}^{\sigma i} \left(n_\sigma \tensor{\partial}_i \alpha\right)  + {\rm h.c.} \right],
\label{eq:L2}
\end{align}
where $\eta_0=\eta_1=\pm1$, with the sign determined respectively by the s- and p-wave effective
ranges. 
$s$ is the spin-singlet auxiliary field of the $nn$ pair, and $\pi_{a}$ 
is the four-component field 
for the  $^2P_{3/2}$ resonance in the $n\alpha$ system. $\tensor{\partial}_i \equiv [(\roarrow{m} \loarrow{\nabla} - \loarrow{m}\roarrow{\nabla})/(\loarrow{m}+\roarrow{m})]_i$ indicates the Galilean invariant derivative, where $\loarrow{m}$ (or $\roarrow{m}$) is the mass of the field operated by $\loarrow{\nabla}$ (or $\roarrow{\nabla}$). Following the convention of Ref.~\cite{Zhang:2013kja}, we define the spin projections of the fields and operators by their indices with $\sigma,\delta,..=\pm1/2$, $a,b,..=\pm 1/2,\pm3/2$, and $i,j,..=0,\pm 1$. $T_{\cdots}^{\cdots}$ is the shorthand notation for the Clebsch-Gordan coefficient, e.g., $T_{a}^{\sigma i} = T^{a}_{\sigma i} \equiv C(1/2, 1, 3/2\,|\, \sigma\, i\, a)$. 

The three-body Lagrangian $\mathscr{L}_3$ describes the $nn\alpha$ contact interaction, whose existence and specific form are derived as a consequence of three-body renormalization (see Sec.~\ref{sec:3body-he6}).
\begin{equation}
\mathscr{L}_3 
= -h \left( T^{ab}_{0} T^{i \sigma}_{a} \pi_b \tensor{\partial}_i n_\sigma \right)^{\dagger} 
\left( T^{cd}_{0} T^{j \delta}_{c} \pi_d \tensor{\partial}_j n_\delta \right)
\label{eq:L3}
\end{equation}

The $nn$ interaction is dominated by an s-wave virtual state
, where the scattering length, $a_0=-18.7$ fm~\cite{GonzalezTrotter:2006wz}, is approximately one order of magnitude larger than the corresponding effective range, $r_0=2.75$ fm~\cite{Slaus:1989}.
According to the
effective-range expansion at low energies, the $nn$  t-matrix for elastic
scattering can be written up to second order in the expansion in powers of $M_{lo}/M_{hi}$ as
\begin{align}
\label{eq:t0-range}
\langle\mathbf{k}|t_{nn}|\mathbf{k'}\rangle = -\frac{1}{4\pi^2 \mu_{nn}}\,\frac{1}{k\cot\delta_0 -ik} =  \frac{1}{4\pi^2 \mu_{nn}}\,\frac{1}{1/a_0 - r_0 k^2/2 +ik}~,
\end{align}
where 
$\mu_{nn}$ is the reduced mass in the $nn$ center-of-mass frame,
$k\equiv |\mathbf{k}|=|\mathbf{k'}|=\sqrt{2\mu_{nn} E}$ indicates the on-shell relative momentum
in the $nn$ subsystem, and $E$ is the 
two-body energy.
Terms with higher  powers of $k$,
such as shape-dependent terms in the effective-range expansion, are suppressed at low momentum.
The free state $|\mathbf{p}\rangle$ is normalized as 
\begin{align}
\label{eq:norm}
\langle\mathbf{p}|\mathbf{p'}\rangle=\delta^{(3)}(\mathbf{p}-\mathbf{p'})~.
\end{align}
The result (\ref{eq:t0-range}) is an exact result for the $nn$ t-matrix, given the Lagrangians (\ref{eq:L1}) and (\ref{eq:L2}). For the relationship between the Lagrangian parameters and the effective-range parameters, see, e.g. Ref.~\cite{Zhang:2013kja}. 

Similarly to the power counting employed in short-range EFT (SREFT) for boson-boson s-wave
scattering, $a_0$, is associated with the low-momentum scale, $a_0\sim 1/ M_{lo}$, while $r_0$ is related to the high-momentum scale of short-distance physics, $r_0\sim1/M_{hi}$. At leading order (LO) in an expansion in powers of $M_{lo}/M_{hi}$, the position of the $nn$ s-wave virtual state (since $a_0<0$) is given by
$\gamma_0=1/a_0$. Hereafter the LO t-matrix for $nn$ scattering is written as
\begin{align}
\label{eq:tnn}
\langle\mathbf{k}|t_{nn}|\mathbf{k'}\rangle = \frac{1}{4\pi^2\mu_{nn} (\gamma_0+ik)}~,
\end{align}
where the $r_0$ dependent term in Eq.~\eqref{eq:t0-range} is dropped at LO.

The $n\alpha$ interaction is dominated by a p-wave resonance. Based on the p-wave effective-range expansion at low energies, the dominant part of the $n\alpha$ t-matrix can be expressed as
\begin{align}
\label{eq:t1-range}
\langle\mathbf{k}|t_{n\alpha}|\mathbf{k'}\rangle = -\frac{3}{4\pi^2 \mu_{n\alpha}}\,\frac{\mathbf{k}\cdot \mathbf{k'}}{k^3\cot\delta_0 -ik^3} = \frac{3}{4\pi^2 \mu_{n\alpha}}\,\frac{\mathbf{k}\cdot \mathbf{k'}}{1/a_1 - r_1 k^2/2 +ik^3},
\end{align}
where $k=\sqrt{2\mu_{n\alpha}E}$, and $\mu_{n\alpha}$ is the reduced mass in the $n\alpha$ center-of-mass frame. Eq.~(\ref{eq:t1-range}) is the full result for the $n \alpha$ t-matrix, given the Lagrangians (\ref{eq:L1}) and (\ref{eq:L2}). The relationship between p-wave effective-range and Lagrangian parameters can be deduced from the results of Ref.~\cite{Hammer:2011ye}.

Here we adopt the power-counting of Bedaque {\it et al.}~\cite{Bedaque:2003wa}, who assumed $1/a_1\sim M_{lo}^2 M_{hi}$ and $r_1 \sim M_{hi}$. Based on this power counting, the $n\alpha$ interaction has a narrow resonance at a low energy of order $M_{lo}$, with the co-existence of a deep bound state $\sim M_{hi}$. We see this by decomposing the denominator of Eq.~\eqref{eq:t1-range} based on its pole expansion:
\begin{align}
\label{eq:p-pole}
\frac{1}{a_1} -\frac{1}{2}r_1 k^2 +ik^3 = \left(\gamma_1 +ik\right) \left(k^2 +i\frac{k_R^2}{\gamma_1}k -k_R^2\right) =0~,
\end{align}
where $\gamma_1$ indicates the position of the bound-state pole, and $k_R$ is the resonant
momentum. The position of the resonance, together with its width, is determined from
Eq.~\eqref{eq:p-pole} as
\begin{align}
k_{\pm} = \pm k_R \sqrt{1-\frac{k_R^2}{4\gamma_1^2}}\, - i\frac{k_R^2}{2\gamma_1}~.
\end{align}

From Eq.~\eqref{eq:p-pole}, we can relate $\gamma_1$ and $k_R$ to $a_1$ and $r_1$ by
\begin{subequations}
\label{eq:a1-r1}
\begin{align}
\frac{1}{a_1} =& -\gamma_1 k_R^2
\\
\frac{r_1}{2} =& \frac{k_R^2}{\gamma_1} -\gamma_1~.
\end{align}
\end{subequations}
Based on the power-counting introduced above, we obtain that $\gamma_1\sim M_{hi}$ and $k_R\sim M_{lo}$. 

The deep bound state $\gamma\sim M_{hi}$ does not affect low-energy physics. Meanwhile, the resonance poles can be rewritten in the $M_{lo}/M_{hi}$ expansion
as 
\begin{align}
k_{\pm} = \pm k_R - i k_R^2/(2\gamma_1) + \mathcal{O}(M_{lo}^3/M_{hi}^2).
\end{align}
The resonance width (imaginary part) is thus one order higher than the resonance position (real part).
 
Unless we happen to be in the vicinity of the resonance, we then obtain, at LO:
\begin{align}
\gamma_1 =& -\frac{r_1}{2}
\\
k_R =& \sqrt{\frac{2}{a_1 r_1}}~.
\end{align}
Therefore, the LO part of the $n\alpha$ scattering t-matrix is expressed as
\begin{align}
\label{eq:tnc}
\langle\mathbf{k}|t_{n\alpha}|\mathbf{k'}\rangle = \frac{3 \mathbf{k}\cdot\mathbf{k'}}{4\pi^2\mu_{n\alpha} \gamma_1(k^2-k_R^2)}~,
\end{align}
where the unitary term $ik^3$ in Eq.~\eqref{eq:t1-range} is treated as a perturbation and is dropped at LO.
Note that the deep bound state does not appear in this LO amplitude: This t-matrix only has 
two poles, at $k=\pm k_R$ on the real $k$-axis, which correspond to the resonance. Since here we are only interested in the bound-state $^6$He, the energy of the $n\alpha$ subsystem must be negative ($k^2<0$). Therefore the singularity in Eq.~\eqref{eq:tnc} does not cause any numerical issues in our calculations.

\subsection{Partial-wave decomposition of the two-body t-matrix}
\label{sec:2body-he6-partial}

In this subsection we explicitly give  the partial wave decomposition of the two-body t-matrix  in
order to establish our conventions. The  momentum space Lippmann-Schwinger equation is given by
\begin{align}
\label{eq:lippman}
 \langle\mathbf{p}|t(E)|\mathbf{p'}\rangle = 
 \langle\mathbf{p}|V|\mathbf{p'}\rangle +\int d^3 q\, \langle\mathbf{p}|V|\mathbf{q}\rangle G_0(q;E) \langle\mathbf{q}|t(E)|\mathbf{p'}\rangle~,
\end{align}
where $\mathbf{p}$ and $\mathbf{p'}$ denote the two-body relative momenta, 
$G_0$ is the free Green's
function in the two-body system.
Defining
partial-wave components of the potential, $v_l(p,p')$, via
\begin{align}
v_l(p,p') \equiv \frac{1}{2} \int^{1}_{-1} \langle\mathbf{p}|V|\mathbf{p'}\rangle\, \textrm{P}_l(\hat{\mathbf{p}}\cdot \hat{\mathbf{p}}')\, d (\hat{\mathbf{p}}\cdot \hat{\mathbf{p}}') ~,
\end{align}
where $\mathrm{P}_l$ is the $l$th Legendre polynomial, 
and analogously for $t_l(p,p';E)$, we obtain
\begin{align}
t_l(p,p';E) = v_l(p,p') + 4\pi \int_0^\infty dq\, q^2 v_l(p,q)\, G_0(q;E)\, t_l(q,p';E)~. 
\end{align}

In our case all two-body interactions have a resonance in one particular partial wave, which
dominates the behavior of the t-matrix. Considering only the dominant part, we have
\begin{align}
\label{eq:tlpp}
\langle\mathbf{p}|t(E)|\mathbf{p'}\rangle = 
(2l+1)\, t_l(p,p';E)\, \textrm{P}_l(\hat{\mathbf{p}}\cdot \hat{\mathbf{p}}')~,
\end{align}
with $l=0$ and $1$ indicating the s- and p-wave two-body interactions. To simplify the
calculation, we will study $t_l(p,p';E)$ using the formalism of separable potentials. Here we
define the $l$th partial wave of the Hermitian two-body potential in a separable form as
\begin{align}
\label{eq:vl-sep}
v_l(p,p') = \lambda_l\, g_l(p)\, g_l(p')~,
\end{align}
where $g_l(q)$ is the form factor, which only depends on the magnitude of $\mathbf{q}$. By
substituting Eq.~\eqref{eq:vl-sep} into Eq.~\eqref{eq:tlpp}, $t_l(p,p';E)$ is then also  separable,
\begin{align}
\label{eq:tl-sep}
t_l(p,p';E) = g_l(p)\, \tau_l(E)\, g_l(p')~,
\end{align}
where the function $\tau_l$ is given as
\begin{align}
\label{eq:taul-int}
\tau_l^{-1} (E) = \frac{1}{\lambda_l} - 4\pi \int^\infty_0 dq\, \frac{q^2}{E-q^2/(2\mu)+i\epsilon} g_l^2(q)~,
\end{align}
and only depends on the energy $E$ of the two-body system.

To reproduce the physical two-body scattering amplitude, the integral in Eq.~\eqref{eq:taul-int}
needs to be regularized and renormalized. For a particular partial wave, the low-energy behavior
of the two-body t-matrix is determined by the effective-range expansion. By choosing a particular
form factor $g_l$ we can regulate the integral in Eq.~\eqref{eq:taul-int} and then tune the
two-body coupling constant $\lambda_l$ to absorb the resulting regularized divergence, thereby
reproducing the parameters in the effective-range expansion. By doing so, the t-matrix is
renormalized, and the dependence of the low-energy physics on the choice of $g_l(q)$ disappears.

\subsection{Two-body renormalization with a separable potential}
\label{sec:2body-he6-renorm}

One regularization method is to introduce Yamaguchi form factor  to describe the two-body interaction,
i.e. writing the form factor as 
\begin{align}
\label{eq:gl-yamaguchi}
g_l(q) = \frac{\beta_l^{2(l+1)}}{(q^2+\beta_l^2)^{l+1}}\,q^l~.
\end{align}
Here $\beta_l$ indicates the high-momentum scale that regularizes the integrals in
Eq.~\eqref{eq:taul-int}. The renormalization of the two-body t-matrix using such a potential is discussed in Appendix~\ref{app:yamaguchi}.
As early as the 1970s, Ghovanlou and Lehman in Ref.~\cite{Ghovanlou:1974zza} used these
form factors to represent the two-body short-distance physics, hoping to determine three-body observables in $^6$He without the input of three-body parameters. However, their value of the $^6$He ground-state binding energy underpredicts the experimental value. The introduction of a three-body force may be a more effective way to obtain an accurate description of the ${}^6$He nucleus using simple two-body potentials. After all, low-energy three-body physics is insensitive to short-distance details of the input two- and three-body interactions.

In this section we introduce a hard cutoff, $\Lambda$, to regularize the ultraviolet divergence in Eq.~\eqref{eq:taul-int},
\begin{align}
\label{eq:gl}
g_l(p) = p^{l}\theta(\Lambda-p)~, 
\end{align}
where $\theta(x)$ denotes the Heaviside step function: $\theta(x)=0$ for $x<0$, and $1$ for $x>0$.

For the s-wave $nn$ interaction ($l=0$) we obtain
\begin{align}
\label{eq:tau0-rg}
\langle \mathbf{k} | t_{nn} | \mathbf{k}' \rangle =& \tau_{nn} (E)
\nn
 =& \left[ \frac{1}{\lambda_0} + 8\pi\mu_{nn} \int^\Lambda_0 dq\, \frac{q^2}{(q^2-k^2-i\epsilon)} \right]^{-1}
\nn
=& \frac{1}{4\pi^2\mu_{nn}} \left[\frac{1}{4\pi^2\mu_{nn} \lambda_0} + \frac{2\Lambda}{\pi} -\frac{2}{\pi\Lambda}k^2 +ik +\mathcal{O}\left(\frac{k^3}{\Lambda^2}\right)\right]^{-1}~.
\end{align}

We can relate $a_0$ and $r_0$ to $\lambda_0$ and $\Lambda$ by
\begin{subequations}
\label{eq:a0r0-pc}
\begin{align}
\label{eq:a0-pc}
\frac{1}{a_0}=&\frac{1}{4\pi^2\mu_{nn} \lambda_0} + \frac{2\Lambda}{\pi}~,
\\
\label{eq:r0-pc}
\frac{r_0}{2} =& \frac{2}{\pi\Lambda}~.
\end{align}
\end{subequations}
By tuning $\lambda_0$ in Eq.~\eqref{eq:a0-pc} to cancel the divergent piece $\sim \Lambda$, we can
obtain $a_0$ of order $1/M_{lo}$. Eq.~\eqref{eq:r0-pc} shows that the condition $r_0\sim 1/M_{hi}$
is naturally maintained if we keep $\Lambda\sim M_{hi}$. But physics is independent of $\Lambda$
if additional higher-order terms are included in  $\mathscr{L}$.
Therefore, we obtain the renormalized $\tau_{nn}$ in the limit $\Lambda\rightarrow\infty$ as
\begin{align}
\label{eq:tau-nn}
\tau_{nn}(E) = \frac{1}{4\pi^2\mu_{nn} (\gamma_0+ik)}~,
\end{align}
where $k=\sqrt{2\mu E}$. This corresponds to the leading-order t-matrix of Eq.~\eqref{eq:tnn}.

In the case of p-wave ($l=1$) $n\alpha$ scattering Eq.~\eqref{eq:gl} leads to a regularized t-matrix for $n\alpha$ scattering,
\begin{align}
\label{eq:t1-rg}
 \langle\mathbf{k}|t_{n\alpha}|\mathbf{k'}\rangle 
 =& 
3 \mathbf{k} \cdot \mathbf{k}' \, \tau_{n\alpha} (E) 
\nn
=& \frac{3\mathbf{k}\cdot\mathbf{k'}}{4\pi^2\mu_{n\alpha} } \left(\frac{1}{4\pi^2\mu_{n\alpha} \lambda_1} + \frac{2\Lambda^3}{3\pi} +\frac{2\Lambda}{\pi}k^2 +ik^3 +\cdots\right)^{-1}~.
\end{align}

In order to renormalize $t_{n\alpha}$ with one fine-tuning, $a_1$ and $r_1$ must satisfy
\begin{subequations}
\label{eq:a1r1-pc}
\begin{align}
\frac{1}{a_1} =&\frac{1}{4\pi^2\mu_{n\alpha} \lambda_1} + \frac{2\Lambda^3}{3\pi}~,
\\
\label{eq:r1-pc}
\frac{r_1}{2} =& -\frac{2\Lambda}{\pi}~.
\end{align}
\end{subequations}
Eq.~\eqref{eq:r1-pc} indicates that $r_1\sim M_{hi}$, which agrees with the power-counting analysis in Ref.~\cite{Bedaque:2003wa}. After tuning $\lambda_1$ to cancel the $\sim \Lambda^3$ divergence, we reproduce $a_1$ to its physical value. Since $a_1\sim 1/(M_{lo}^2 M_{hi})$, it has the same order as the $r_1 k^2/2$ term in the effective-range expansion. Therefore, the p-wave effective-range parameter $r_1$ must be included at leading order, which agrees with our previous analysis in Subsection \ref{sec:2body-he6-lo}.

Thus, after renormalization, we find for $\tau_{n\alpha}$
\begin{align}
\label{eq:tau-nc}
\tau_{n\alpha} =\frac{1}{4\pi^2\mu_{n\alpha} \gamma_1(k^2-k_R^2)}~.
\end{align}

In contradistinction, the power counting of Ref.~\cite{Bertulani:2002sz} with $a_1\sim M_{lo}^{-3}$ and $r_1\sim M_{lo}$ requires two fine tunings in Eq.~\eqref{eq:t1-rg} to renormalize $t_{n\alpha}$, and yields a different LO expression for $\tau_{n\alpha}$.


\section{Spin and angular momenta in the $^6$He ground state}
\label{sec:Jacobi-he6}

The ground-state of $^6$He has total angular momentum and parity $J=0^+$. Its two-neutron separation
 energy is $0.975$ MeV. 
In this paper we will use Jacobi-momenta $\boldsymbol{K}$, $\boldsymbol{q}_i$, and $\boldsymbol{p}_i$ to represent the internal
kinematics of the three-body system.  Here $\boldsymbol{K}$ is the total momentum, which is zero in the
center-of-mass frame, and $\boldsymbol{q}_i$ and $\boldsymbol{p}_i$ are the relative momenta. The index $i$ on the relative momenta indicates
that they are defined in    
the two-body fragmentation channel $(i,jk)$,  
in which particle $i$ is the spectator and  $(jk)$ the interacting pair. 
Based on this definition, $\boldsymbol{p}_i$ indicates the relative momentum in the $(jk)$ pair, while $\boldsymbol{q}_i$ denotes the relative momentum between the spectator $i$ and the $(jk)$ pair.
Plane-wave states are normalized according to:
\begin{align}
\int d\boldsymbol{p}_i d\boldsymbol{q}_i d\boldsymbol{K} |\boldsymbol{p}_i \boldsymbol{q}_i \boldsymbol{K}\rangle \langle
\boldsymbol{p}_i \boldsymbol{q}_i \boldsymbol{K}| = \mathds{1}.
\end{align}
We define the relative orbital angular momentum and the spin in the pair $(jk)$ as $l_i$ and $s_i$, and the relative total angular momentum for this pair in spin--and--orbital-angular-momentum coupling as $j_i$. In this representation, we also define the relative orbital angular momentum and spin between the spectator $i$ and the pair $(jk)$ as $\lambda_i$ and $\sigma_i$, and the corresponding total angular momentum as $I_i$. Furthermore, the overall orbital angular momentum, spin and total angular momentum of the three-body system are defined as $L_i$, $S_i$ and $J$. Due to spin and angular-momentum conservation, these quantum numbers must obey
\begin{subequations}
\begin{align}
\mathbf{L}_i =& \mathbf{l}_i+\mathbf{\lambda}_i~,
\\
\mathbf{S}_i =& \mathbf{s}_i+\mathbf{\sigma}_i~,
\\
\mathbf{J} =& \mathbf{L}_i+\mathbf{S}_i = \mathbf{j}_i+\mathbf{I}_i~.
\end{align}
\end{subequations}
With the $\alpha$-core as the spectator, we obtain $l_\alpha=s_\alpha=j_\alpha=0$, since the $nn$ interaction is dominated by the $^1$S$_0$ virtual state.
Furthermore, at LO $\lambda_\alpha=\sigma_\alpha=I_\alpha=0$ and it is then straightforward to determine that $S_\alpha=L_\alpha=0$ in the $(\alpha, nn)$ partition. 
Alternatively, if we choose a neutron as the spectator, the $n\alpha$ interaction is dominated by the $^2$P$_{3/2}$ resonance, which means $l_n=1$, $s_n=1/2$ and $j_n=3/2$. Therefore, in the $^6$He ground state, the spectator neutron must also interact with the $n\alpha$ pair in a p-wave, because of the positive parity of the $^6$He ground state. This results in $\lambda_n=1$, $\sigma_n=1/2$. Since $\mathbf{j}_n+\mathbf{I}_n=\mathbf{J}=0$, we must have $I_n=3/2$. In the 
$(n, n\alpha)$ partition, the spin-spin and orbit-orbit couplings have two possibilities: the overall orbital angular momentum and overall spin can either be both zero ($L_n=S_n=0$) or both $1$ ($L_n=S_n=1$). These two cases can contribute to the $^6$He $J=0^+$ state. We summarize the possible spin and orbital-angular-momentum properties of the $^6$He ground state in Table~\ref{table:spin-angular} with respect to different spectator partitions.

\begin{table}
\caption{Spin and orbital-angular-momentum coupling in $^6$He to obtain its ground state $J=0^+$.}
\begin{center}
\renewcommand{\arraystretch}{1.5}
\begin{tabularx}{0.7\linewidth}{|*{5}{>{\centering\arraybackslash}X|}}
\cline{1-5}
(spectator,pair) & pair	& spectator & total $L$, $S$ & total $J^{P}$ \\
\cline{1-5}
( $\alpha$,$nn$)	& $\ell_\alpha=0$, $s_\alpha=0$ & $\lambda_\alpha=0$, $\sigma_\alpha=0$ &
$L_\alpha=0$, $S_\alpha=0$ & \multirow{2}{*}{$J=0^{+}$} \\
\cline{1-4}
\multirow{2}{*}{($n$,$n\alpha$)}	& \multirow{2}{*}{$\ell_n=1$, $s_n=\frac{1}{2}$} &
\multirow{2}{*}{$\lambda_n=1$, $\sigma_n=\frac{1}{2}$} & $L_n=0$, $S_n=0$	&
\\
\cline{4-4}
	&		&	& 
 $L_n=1$, $S_n=1$ 	&
\\
\cline{1-5}
\end{tabularx}
\end{center}
\label{table:spin-angular}
\end{table}

Knowing the spin and orbital-angular-momentum quantum numbers, we can construct an eigenstate of $^6$He with respect to the spin and orbital-angular-momentum operators. Considering all conserved quantities in the three-body system, we decompose the Jacobi momenta 
with respect to these spin and orbital- and total-angular-momentum quantum numbers by
\begin{equation}
\label{eq:quantum-number}
\left|p,q;\Omega_i\right\rangle_{i} = \sum\limits_{L_i S_i}\sqrt{\widehat{j}_i \widehat{I}_i \widehat{L}_i \widehat{S}_i}\,
\begin{Bmatrix}
  l_i & s_i &  j_i \\

  \lambda_i & \sigma_i &  I_i \\
  L_i     & S_i     & J 
\end{Bmatrix}\,
\left| p,q;(l_i,\lambda_i)L_i\,; \left[(\nu_j \nu_k)s_i,\sigma_i\right]S_i\,;J=M_J=0\right\rangle_{i}~,
\end{equation}
where $\widehat{j}_i$ denotes $2{j}_i+1$ (the same holds for $\widehat{I}_i$, $\widehat{L}_i$ and $\widehat{S}_i$), $p\equiv|\mathbf{p}|$, and $q\equiv|\mathbf{q}|$. Meanwhile $\Omega_i$ represents all conserved spin, orbital- and total-angular-momentum quantum numbers in the  partition  $(i, jk)$. Those quantum numbers are included in the Wigner 9j symbol in Eq.~\eqref{eq:quantum-number}. In addition, the labels  $\nu_j$ and $\nu_k$ in the bracket denote the individual spins of particle $j$ and $k$. They are coupled to produce the spin $s_i$ of the pair $(jk)$.

Applying Eq.~\eqref{eq:quantum-number} in the partition $(\alpha, nn)$ the eigenstate of the $^6$He ground state can be written as
\begin{align}
\left|p,q;\Omega_\alpha\right\rangle_{\alpha} =\left| p,q;\,(0,0)L_\alpha=0;\, \left[\left(\frac{1}{2}\frac{1}{2}\right)0,0\right]S_\alpha=0;\,J=M_J=0\right\rangle_{\alpha}~.
\end{align}
Similarly, in the partition $(n, n\alpha)$, the $^6$He ground-state eigenstate can be expressed as
\begin{align}
\left|p,q;\Omega_n\right\rangle_{n} =\sum\limits_{L_n=0}^{1} \sqrt{\frac{2}{3}} \left(\frac{-1}{\sqrt{2}}\right)^{L_n} \left| p,q;\,(1,1)L_n;\, \left[\left(\frac{1}{2}0\right)\frac{1}{2},\frac{1}{2}\right]S_n=L_n;\,J=M_J=0 \right\rangle_{n}~.
\end{align}

We can further decouple the orbital angular momentum and the spin by using the Clebsch-Gordan coefficient $C(LSJ|M_L M_S M_J)$. In the 
$(\alpha, nn)$ basis we obtain
\begin{align}
\label{eq:state-alpha-LS}
|p,q;\Omega_\alpha \rangle_{\alpha} = \left| p,q;\,0,0; L_\alpha=0, M_{L\alpha}=0 \right\rangle_{\alpha}\, \left|\left(\frac{1}{2}\, \frac{1}{2}\right) \,0,0;S_\alpha=0, M_{S\alpha}=0\right\rangle_{\alpha}~,
\end{align}
while in the $(n, n\alpha)$ basis, we find
\begin{align}
\label{eq:state-n-LS}
|p,q;\Omega_n \rangle_{n} =& \sum\limits_{L_n=0}^{1} \sum\limits_{M_{Ln}=-L_n}^{L_n} (-1)^{M_{Ln}} \sqrt{\frac{2^{1-L_n}}{6L+3}}
\left| p,q;\,1,1;\,L_n, M_{Ln}\right\rangle_{n}\,
\left|\left(\frac{1}{2}0\right)\frac{1}{2},\frac{1}{2};\,S_n=L_n,M_{Sn}=-M_{Ln}\right\rangle_{n}~.
\end{align}

\section{Halo EFT in the three-body sector}
\label{sec:3body-he6}

In this section, we study the behavior of $^6$He as a three-body problem in halo EFT. We focus on the three-body bound-state problem, and set up the Faddeev equations, based on Refs.~\cite{Glockle:1983, Afnan:1977}, for solving 
for the three-body binding energy of $^6$He.
We then employ the formalism to investigate the ground state of $^6$He projected on to the particular partial waves discussed in Sec.~\ref{sec:Jacobi-he6}. Without a $nn\alpha$ three-body counterterm, the results will be cutoff dependent. Therefore, we need to discuss the regularization and renormalization procedures in our analysis. By adding a $nn\alpha$ counterterm, we reproduce the experimental value of the $^6$He two-neutron separation energy, $S_{2n}=0.975$ MeV, and predict the  Faddeev components.

\subsection{Faddeev decomposition of the three-body wave function}
Considering only two-body potentials, the  general Schr\"{o}dinger equation in a 
system with three distinguishable particles reads
\begin{align}
\label{eq:schrodinger}
\left(H_0 + \sum\limits_{i=1}^3 V_i\right) |\Psi\rangle =E\,|\Psi\rangle~,
\end{align}
where $V_i$ indicates the potential between particles $j$ and $k$ in the partition $(i,jk)$. Following Faddeev~\cite{Faddeev:1960su}, the
wave function is decomposed into three components, one with respect to each of the three different spectators,
\begin{align}
\label{eq:Psi-i}
|\Psi\rangle = \sum\limits_{i=1}^3 |\psi_i\rangle~,
\end{align}
with
$|\psi_i\rangle$ being the Faddeev component in the $(i,jk)$ partition.
Inserting Eq.~\eqref{eq:Psi-i} into Eq.~\eqref{eq:schrodinger} and employing the Lippmann-Schwinger equation we 
obtain~\cite{Glockle:1983}
\begin{align}
\label{eq:faddeev-psi}
|\psi_i\rangle = G_0 t_i \sum\limits_{j\neq i} |\psi_j\rangle~.
\end{align}
Here $t_i$ represents the two-body t-matrix for the pair $(jk)$, $t_i\equiv t_{jk}$. 
All components are obtained by a cyclic permutation of $(i,jk)$. Note that Eq.~\eqref{eq:faddeev-psi} is a homogeneous integral equation, since only the bound state is considered.

To simplify our future calculations, we define new components $|F_i\rangle$, which are related to  $|\psi_i\rangle$ by
\begin{align}
\label{eq:Fi-psi}
|\psi_i\rangle= G_0 t_i |F_i\rangle~.
\end{align}
By substituting Eq.~\eqref{eq:Fi-psi} into Eq.~\eqref{eq:faddeev-psi}, we obtain the Faddeev equation for $|F_i\rangle$:
\begin{align}
\label{eq:faddeev-Fi}
|F_i\rangle = \sum\limits_{j\neq i} G_0 t_j |F_j\rangle~.
\end{align}

If the two-body t-matrix, $t_i$, is separable, 
then its matrix presentation in the basis of eigenstates $\{\, | p,q,;\Omega_i\rangle_{i} \} $
leads to a relatively simple expression in which the momenta $p$, $p'$, and $q$ are decoupled,
\begin{align}
\label{eq:ti-project}
\,_{i}\langle p,q;\Omega_i| t_i|p',q';\Omega_i'\rangle_{i} = 4\pi\, g_{l_i}(p)\, \tau_i(q;E)\, g_{l_i}(p')\, \delta_{\Omega_i \Omega_i'}\, \frac{1}{q^2}\delta(q-q')~, 
\end{align}
provided that the two-body t-matrix is diagonal in the quantum numbers $\Omega_i$. 
In our case $t_i$ operates only in a specific partial wave: ${}^1$S$_0$ for $nn$ and ${}^2$P$_{3/2}$ for $n\alpha$.  Eq.~\eqref{eq:ti-project} only gives the two-body t-matrix's matrix elements in three-body Hilbert space when $\Omega_i$ corresponds to those particular two-body channels. The matrix elements in all other channels are zero in our LO calculation.
The quantity $\tau_i$ in Eq.~\eqref{eq:ti-project} is related to $\tau_{jk}$ of Eqs.~(\ref{eq:tau-nn}) and (\ref{eq:tau-nc}) by
\begin{align}
\tau_i(q;E) \equiv \tau_{jk}\left(E-\frac{q^2}{2 \mu_{i(jk)}}\right)~.
\end{align}
Here $E$ denotes the total energy of the three-body system relative to the $\alpha nn$ threshold, and $\mu_{i(jk)}$ is the reduced mass with respect to the spectator $i$ and the pair $(jk)$. We are interested in $E=-B_3$, with $B_3 > 0$ the binding energy of the three-body system, which, for two-neutron halos, is the two-neutron separation energy of the nucleus, i.e., $B_3=S_{2n}$.

Projecting  Eq.~(\ref{eq:faddeev-Fi}) on to the state $|p,q;\Omega_i\rangle_{i}$ leads to
\begin{align}
\label{eq:faddeev2}
\,_{i}\langle p, q; \Omega_i |F_i \rangle =& 4\pi \sum\limits_{j\neq i} \iint p'^2 d p'\, q'^2 dq'\, G_0^{(i)}(p,q;E)\,
\,_{i}\langle p, q; \Omega_i |p', q'; \Omega_j\rangle_{j}
g_{l_j}(p')\, \tau_j(q';E)\, \nonumber\\
&\times\int p''^2 d p'' \, g_{l_j}(p'')\,_{j}\langle p'', q'; \Omega_j | F_j\rangle~,
\end{align}
where $G_0^{(i)}$ is the momentum representation of the three-body Green's function with respect to the spectator $i$:
\begin{align}
\label{eq:G0-i}
G_0^{(i)}(p,q;E) = \left(E-\frac{p^2}{2\mu_{jk}} - \frac{q^2}{2\mu_{i\,(jk)}}\right)^{-1}~.
\end{align}

Absorbing the dependence on the inter-pair momentum, $p$, in the Faddeev equation~\eqref{eq:faddeev2}, we can construct a simplified integral equation in which quantities depend only on the relative momentum between the spectator and the pair, $q$. To achieve this, we define a new function $F_i(q)$,
\begin{align}
\label{eq:Fi-q}
F_i(q) = \int p^2 dp\, g_{l_i}(p)\, \,_{i}\langle p, q; \Omega_i |F_i \rangle~.
\end{align}
By substituting Eq.~\eqref{eq:Fi-q} into Eq.~\eqref{eq:faddeev2}, we find that 
\begin{align}
\label{eq:faddeev1}
F_i(q) = \sum\limits_{j\neq i} 4\pi \int q'^2 dq'\, X_{ij}(q,q';E)\, \tau_j(q';E)\, F_j(q')~.
\end{align}
The kernel function $X_{i j}$ is defined by
\begin{align}
\label{eq:faddeev3}
X_{i j}(q,q';E) = \iint p^2 dp\, p'^2 d p'\, g_{l_i}(p) G_0^{(i)}(p,q;E)\, g_{l_j}(p')\, \,_{i}\langle p, q; \Omega_i |p', q'; \Omega_j\rangle_{j}~,
\end{align}
which includes the three-body Green's function $G_0^{(i)}$ and the two-body form factors $g_{l_i}$ and $g_{l_j}$. The factor $\,_{i}\langle p, q; \Omega_i |p', q'; \Omega_j\rangle_{j}$ is the projection of the eigenstate of the free Hamiltonian in the partition of spectator $i$ onto the
free eigenstate in the partition of spectator $j$~\cite{Glockle:1983}.

To solve this integral equation~\eqref{eq:faddeev1}, we look for an energy $E=-B_3$ where the eigenvalue of the kernel is one.

\subsection{Faddeev equations for the $^6$He system}

Here we apply the Faddeev formalism established in the previous subsection to the $^6$He ground state. For this purpose Eq.~\eqref{eq:Psi-i} can be re-expressed as
\begin{align}
\label{eq:Psi-ann}
|\Psi\rangle = |\psi_\alpha\rangle + |\psi_n\rangle + |\psi_{n'}\rangle= |\psi_\alpha\rangle +(1-\mathcal{P}_{nn})|\psi_n\rangle~,
\end{align}
where the three terms are the Faddeev components for $(\alpha, nn)$, and the two $(n,n\alpha)$ partitions, with those last two related by fermionic symmetry. Because
the two neutrons are fermions the 
$^6$He wave function must be anti-symmetric under their permutation $\mathcal{P}_{nn}$, and Eq.~(\ref{eq:Psi-ann}) indeed fulfils
\begin{align}
\mathcal{P}_{nn} |\Psi\rangle =  -|\Psi\rangle,
\end{align}
since
\begin{align}
\mathcal{P}_{nn}|\psi_\alpha\rangle=&-|\psi_\alpha\rangle.
\end{align}
The Green's function, $G_0$, and the two-body t-matrices  $t_\alpha$ and $t_n$ are unchanged under the action of $\mathcal{P}_{nn}$, because they were defined above as projections of only the neutron-spin-independent part of the eigenstate.

\begin{figure}[ht]
\centerline{\includegraphics[width=14cm,angle=0,clip=true]{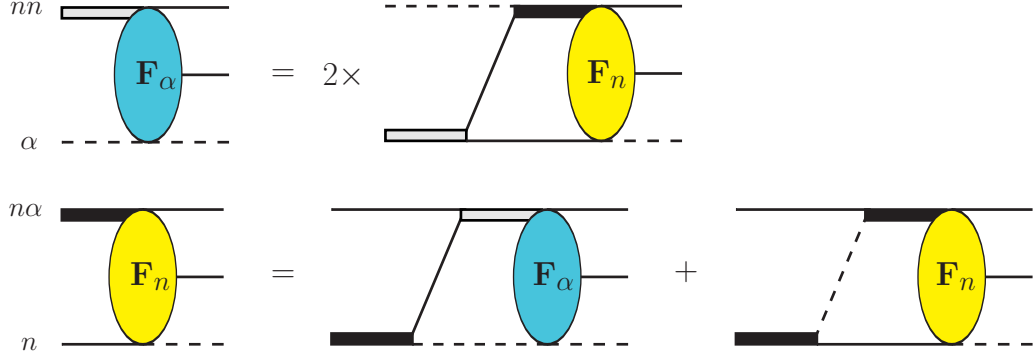}}
\caption{(Color online) The Faddeev equations for the $^6$He bound-state problem: The single dashed (or solid) line denotes the $\alpha$ (or $n$) one-body propagator. The thick black and shaded lines
 represents the $n\alpha$ and $nn$ two-body propagators. 
The ellipses labeled ``$F_\alpha$'' and ``$F_n$'' are the corresponding Faddeev components.}
\label{pic:couple-eq}
\end{figure}

By projecting the Faddeev components $|F_\alpha\rangle$ and $|F_n\rangle$ onto the partial-wave-decomposed states 
in respective partitions we obtain two coupled-channel integral equations for the $^6$He ground state,
\begin{subequations}
\begin{align}
\label{eq:Fc-faddeev}
F_{\alpha}(q) =& 8\pi \int^\Lambda_0 q'^2 dq'\, X_{\alpha n}(q,q';-B_3)\,  \tau_n(q';-B_3)\, F_{n}(q')~;
\\
\label{eq:Fn-faddeev}
F_{n}(q) =& 4\pi \int^\Lambda_0 q'^2 dq'\, X_{n \alpha}(q,q';-B_3)\, \tau_\alpha(q';-B_3)\, F_{\alpha}(q')
+ 4\pi \int^\Lambda_0 q'^2 dq'\, X_{nn}(q,q';-B_3)\, \tau_n(q';-B_3)\, F_n(q')~,
\end{align}
\end{subequations}
where the ultraviolet cutoff, $\Lambda$, is introduced for regularization. 
The Faddeev equations~(\ref{eq:Fc-faddeev}, \ref{eq:Fn-faddeev}) 
are diagrammatically expressed in Fig.~\ref{pic:couple-eq}.
Similar coupled-channel integral equations for two-neutron s-wave halo nuclei were derived by Canham and Hammer in Ref.~\cite{Canham:2008jd}, where the two sets of equations differ only in their expressions for the kernel functions $X_{ij}$. 

The quantities $\tau_\alpha$ and $\tau_n$ appearing in Eqs.~(\ref{eq:Fc-faddeev}) and (\ref{eq:Fn-faddeev}) are functions of the $^6$He two-neutron separation energy $B_3$ and the Jacobi momentum $q$:
\begin{align}
\tau_{\alpha}(q;-B_3) = \frac{1}{2\pi^2 m_n} \frac{1}{\gamma_0 -\mathcal{K}_\alpha(q;-B_3)}~,
\end{align}
where the two-body binding momentum $\mathcal{K}_\alpha$ is related to $q$ and $B_3$ by
\begin{align}
\mathcal{K}_{\alpha}(q;-B_3) = \sqrt{m_n B_3 + \frac{A+2}{4A} q^2}~.
\end{align}
Here $A$ indicates the mass ratio between the $\alpha$-core and a neutron $A= m_\alpha/m_n$.

Similarly, we can  write $\tau_{n}$ as a function of $B_3$ and the Jacobi momentum $q$:
\begin{align}
\label{eq:taun}
\tau_{n}(q;-B_3) = -\frac{1}{4\pi^2 m_n \gamma_1}\left(\frac{A+1}{A}\right)\frac{1}{\mathcal{K}_n^2(q;-B_3) +k_R^2}~,
\end{align}
where $\mathcal{K}_n$ is given as
\begin{align}
\mathcal{K}_n(q;-B_3) = \sqrt{\frac{2A}{A+1}\left(m_n B_3 + \frac{A+2}{2(A+1)}q^2\right)}~.
\end{align}

Meanwhile the kernel functions $X_{n\alpha}$, $X_{\alpha n}$ and $X_{nn}$ are calculated according to Eq.~\eqref{eq:faddeev3}, where the subscripts indicate 
the two spectator partitions involved in the transition.
The details of these calculations are presented in Appendix~\ref{app:kernel}, and result
in the following final expressions:
\begin{subequations}
\label{eq:faddeev-X}
\begin{align}
X_{n\alpha}(q,q';-B_3) 
=& -\sqrt{2}\, m_n \left[\frac{A}{A+1}\, \frac{1}{q'}\, \textrm{Q}_0 (z_{n\alpha}) + \frac{1}{q}\, \textrm{Q}_1 (z_{n\alpha})\right]~,
\\
X_{\alpha n}(q,q';-B_3) 
=& -\sqrt{2} m_n \left[\frac{A}{A+1}\, \frac{1}{q}\, \textrm{Q}_0 (z_{\alpha n}) + \frac{1}{q'}\, \textrm{Q}_1 (z_{\alpha n})\right]~,
\\
X_{nn}(q,q';-B_3) 
=& A\, m_n \left[\frac{A^2+2A+3}{(A+1)^2}\,\textrm{Q}_0(z_{n n})
+\frac{2}{A+1} \frac{q^2+q'^2}{qq'}\,\textrm{Q}_1(z_{n n})
+ \textrm{Q}_2(z_{n n})\right]~,
\end{align}
\end{subequations}
where  $\textrm{Q}_l$ are the Legendre functions of the second kind,
which are related the ordinary Legendre polynomials $\textrm{P}_l$ by
\begin{align}
\label{eq:Ql}
\textrm{Q}_l(z) = \frac{1}{2} \int^{1}_{-1}dx\, \frac{\textrm{P}_l(x)}{z-x}~
\end{align} 
for $|z|>1$. 
The arguments $z_{n\alpha}$, $z_{\alpha n}$ and $z_{n n}$ in Eq.~\eqref{eq:faddeev-X} are defined as
\begin{subequations}
\begin{align}
z_{n\alpha} =& -\frac{1}{qq'}\left(m_n B_3 + q^2 +\frac{A+1}{2A}q'^2\right)~,
\\
z_{\alpha n} =& -\frac{1}{qq'}\left(m_n B_3 + \frac{A+1}{2A} q^2 +q'^2\right)~,
\\
z_{n n} =& -\frac{A}{qq'} \left(m_n B_3 + \frac{A+1}{2A}(q^2+q'^2)\right)~.
\end{align}
\end{subequations}
In our bound-state situation, 
$B_3>0$,
these three arguments all satisfy the condition $z<-1$.

\begin{figure}[ht]
\centerline{\includegraphics[width=14cm,angle=0,clip=true]{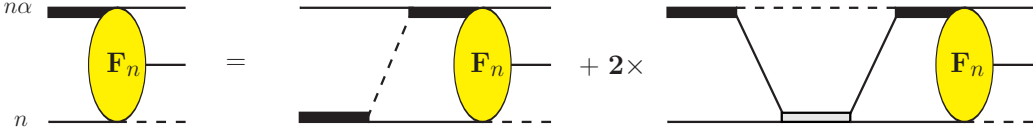}}
\caption{(Color online) The single-channel Faddeev equation for the $^6$He bound state. The Faddeev component $F_n$ is solved as integral equation containing a two-loop diagram.}
\label{pic:single-eq}
\end{figure}

By inserting Eq.~\eqref{eq:Fc-faddeev} into~\eqref{eq:Fn-faddeev}, we obtain a single-channel integral equation that includes only the Faddeev component $F_{n}$: 
\begin{align}
\label{eq:single-Fn}
F_{n}(q) =& 4\pi \int^\Lambda_0 q'^2 dq'\, X_{nn}(q,q';-B_3)\, \tau_n(q';-B_3)\, F_n(q')
\nn
&+8\pi \int^\Lambda_0 q'^2 dq'\, \left[4\pi \int^\Lambda_0 q''^2 dq''\, X_{n \alpha}(q,q'';-B_3)\, \tau_\alpha(q'';-B_3)X_{\alpha n}(q'',q';-B_3)\right]
\tau_n(q';-B_3)\, F_n(q')~.
\end{align}
Eq.~\eqref{eq:single-Fn} can be diagrammatically expressed in Fig.~\ref{pic:single-eq}. The last term in Fig.~\ref{pic:single-eq} contains two loops, which corresponds to the double integral in Eq.~\eqref{eq:single-Fn}. For future reference we define the integral inside the square brackets in Eq.~(\ref{eq:single-Fn})  as the function
\begin{align}
\label{eq:Incn}
I_{n\alpha n}(q,q';B_3)=4\pi \int^\Lambda_0 q''^2 dq''\, X_{n \alpha}(q,q'';-B_3)\, \tau_\alpha(q'';-B_3)X_{\alpha n}(q'',q';-B_3)~.
\end{align}

Eq.~(\ref{eq:single-Fn}) foregrounds a possible inconsistency in our approach. In the Halo EFT power counting both $X_{nn}$ and $I_{n \alpha n}$ are of order $Q^0$. Meanwhile, in the power counting of Ref.~\cite{Bedaque:2003wa} the propagator $\tau_n$ scales as $M_{hi}^{-1} Q^{-2}$ [see Eq.~(\ref{eq:taun})]. It follows that each iterate of the integral equation (term in the Neumann series~\cite{Arfken}) is suppressed by one power of $Q/M_{hi}$ compared to the previous one. If  the theory is properly renormalized in the three-body sector, i.e. only momenta of order $M_{lo}$ contribute to the loop integrations,  our power counting then leads to the conclusion that there are no ${}^6$He bound states. An alternative way to state this is that the power counting of Ref.~\cite{Bedaque:2003wa} predicts that the eigenvalues of the integral-equation kernel are of order $M_{lo}/M_{hi}$, and so there are no solutions to Eq.~(\ref{eq:single-Fn})---provided it is properly renormalized.

Clearly this conclusion is not correct, since ${}^6$He exists.  The power counting of Ref.~\cite{Bertulani:2002sz}, which is less ``natural" in the $n\alpha$ sector (see Sec.~\ref{sec:concl-he6}), does not produce this dilemma in the three-body sector. In that power counting $\tau_n \sim M_{lo}^{-1} Q^{-2}$, and all terms in the Neumann series are of the same size for $Q \sim M_{lo}$. But, in the power counting of Ref.~\cite{Bertulani:2002sz}, Eq.~(\ref{eq:taun}) must also be modified, since the unitarity piece of the $n \alpha$ amplitude is present already  at leading order. The corresponding calculation for ${}^6$He was carried out in Ref.~\cite{Rotureau:2012yu}.

\subsection{Renormalization of the $^6$He ground state}

The conclusion of perturbativity also rests on the assumption that Eq.~(\ref{eq:single-Fn}) is already renormalized. We now show that this is not the case.

\begin{figure}[ht]
\centerline{\includegraphics[width=12cm,angle=0,clip=true]{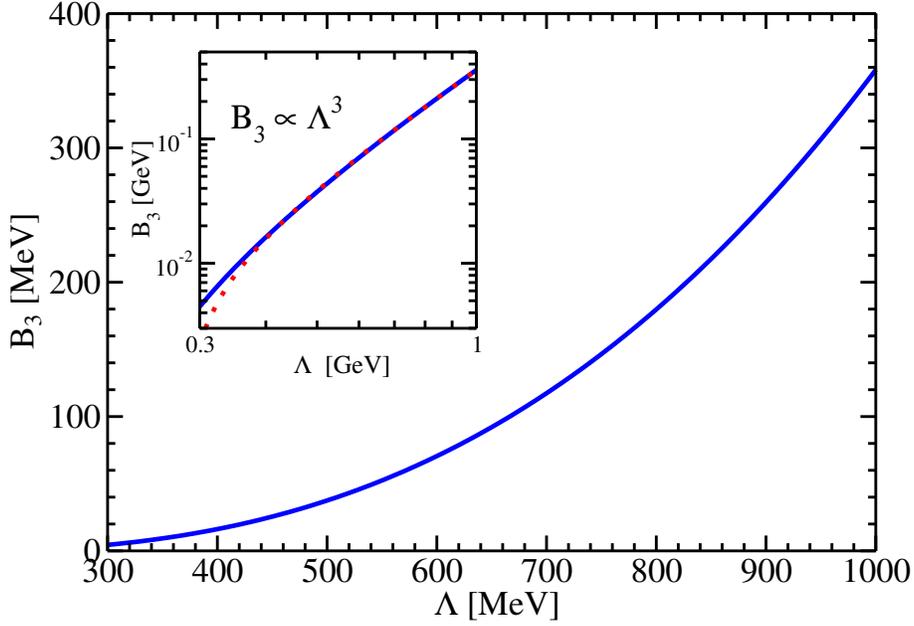}}
\caption{(Color online) The $^6$He two-neutron separation energy $B_3$ as a function of the cutoff $\Lambda$. The calculation  is based on only two-body interactions. The inner panel compares the numerical result (blue solid line) with a polynomial approximation (red dotted line): $B_3/{\rm GeV} = -0.00765+0.366 (\Lambda/{\rm GeV})^3$.}
\label{pic:B3}
\end{figure}

By using a hard cutoff $\Lambda$ to regularize the integrals of Eqs.~(\ref{eq:Fc-faddeev}, \ref{eq:Fn-faddeev}), we obtain $B_3$ as a function of $\Lambda$. This cutoff dependence is illustrated in Fig.~\ref{pic:B3}, which shows that $B_3$ behaves approximately as $\Lambda^{3}$ at values of $\Lambda$ that are large compared to $k_R$, $\gamma_0$, and $\sqrt{2 m_n B_3}$. 

To  understand this phenomenon we examine the properties of the kernel of Eq.~(\ref{eq:single-Fn}).  Since the analytic form of each term in Eq.~\eqref{eq:Incn} is already derived, we can calculate the cutoff dependence of 
$I_{n\alpha n}$ analytically. In fact, the dominant $\Lambda$-dependent part of $I_{n\alpha n}$ is proportional to $ m_n qq'/\Lambda^2$, and vanishes in the limit $\Lambda\rightarrow \infty$. Since $X_{nn}$ is not cutoff dependent, the kernel of the single-channel integral equation, Eq.~\eqref{eq:single-Fn}, is independent of $\Lambda$ for $\Lambda \gg \sqrt{2 m_n B_3}, \gamma_0, k_R$. Thus, the cutoff dependence that appears in Fig.~\ref{pic:B3} must arise from the solution of the integral equation. 

\begin{figure}[ht]
\centerline{\includegraphics[width=14cm,angle=0,clip=true]{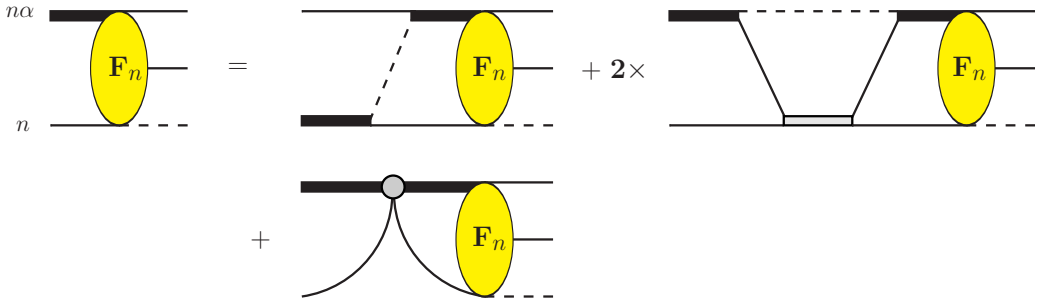}}
\caption{(Color online) The single-channel Faddeev equation for the $^6$He bound state with the addition of a $nn\alpha$ counterterm. Both sides of the counterterm are in the spectator-$n$ partition.}
\label{pic:single-3bf}
\end{figure}

In order to cancel this cutoff dependence, a three-body $nn\alpha$ counterterm is added to the integral equation. A natural choice of this counterterm is one that has the same behavior as the cutoff-dependent piece of $I_{n\alpha n}$, i.e. proportional to $ m_n qq'/\Lambda^2$. The dependence on both $q$ and $q'$ indicates the existence of p-wave channels on both sides of the counterterm. Therefore, we introduce a $nn\alpha$ counterterm with a neutron as the spectator on both sides of the counterterm. 
Choosing a $nn\alpha$ counterterm of this p-wave type is also consistent with the Pauli exclusion principle.
The resulting integral equation with the addition of a $nn\alpha$ counterterm is diagrammatically illustrated in Fig.~\ref{pic:single-3bf}.

In order to include this $nn\alpha$ counterterm, Eq.~\eqref{eq:single-Fn}  needs to be modified by adding the following term to the kernel function ${X}_{nn}$,
\begin{align}
\label{eq:Xnn-3bf}
X_{nn}(q,q';-B_3) \rightarrow  X_{nn}(q,q';-B_3) - m_n \frac{q q'}{\Lambda^2}H_0(\Lambda)~,
\end{align}
where the minus sign in Eq.~\eqref{eq:Xnn-3bf} is introduced due to the presence of the permutation operator $-\mathcal{P}_{nn}$ in the kernel function $X_{nn}$. Since $H_0(\Lambda)$ is itself unchanged under the permutation, applying $-\mathcal{P}_{nn}$ to the three-body force will lead to 
a factor of $-1$.

\begin{figure}[ht]
\centerline{\includegraphics[width=12cm,angle=0,clip=true]{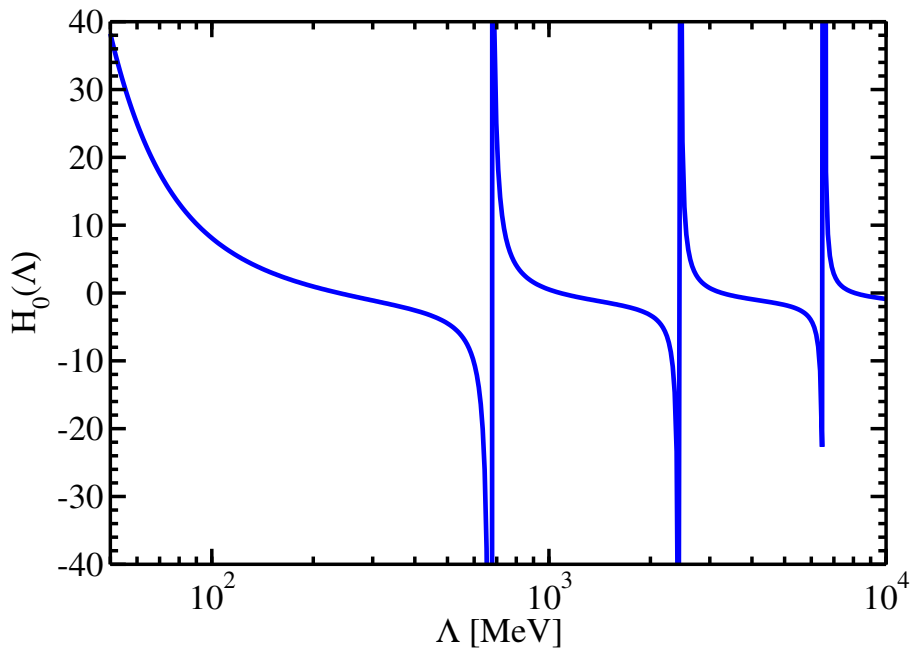}}
\caption{(Color online) The $nn\alpha$-counterterm parameter $H_0$ as a function of the cutoff $\Lambda$. $H_0$ is tuned to reproduce $B_3=0.975$ MeV at different values of $\Lambda$.}
\label{pic:H0-he6}
\end{figure}

By tuning the counterterm parameter $H_0(\Lambda)$, we can cancel the cutoff dependence in the integral equation, Eq.~\eqref{eq:single-Fn}, and reproduce the $^6$He ground-state two-neutron separation energy $B_3 = 0.975$ MeV for all values of $\Lambda$. In Fig.~\ref{pic:H0-he6} we plot the $H_0(\Lambda)$ that is necessary to do this as a function of $\Lambda$. It has an oscillatory behavior in $\log \Lambda$---similar to the three-body force's behavior in the leading-order three-boson problem~\cite{Bedaque:1998kg}. However, in contrast to the three-boson case, the period of $H_0(\Lambda)$ in $\log \Lambda$ decreases as $\Lambda$ increases. This difference in the behavior of $H_0$ may well arise from the $n\alpha$ p-wave interaction in the $^6$He system: The symmetry of discrete scale invariance, present in three-body systems with resonant s-wave interactions, is broken by this p-wave interaction (cf.~\cite{Braaten:2011vf} which considers a three-body system with all p-wave interactions, and in a zero-range 
limit that differs from that discussed 
here).

\begin{figure}[ht]
\centerline{\includegraphics[width=12cm,angle=0,clip=true]{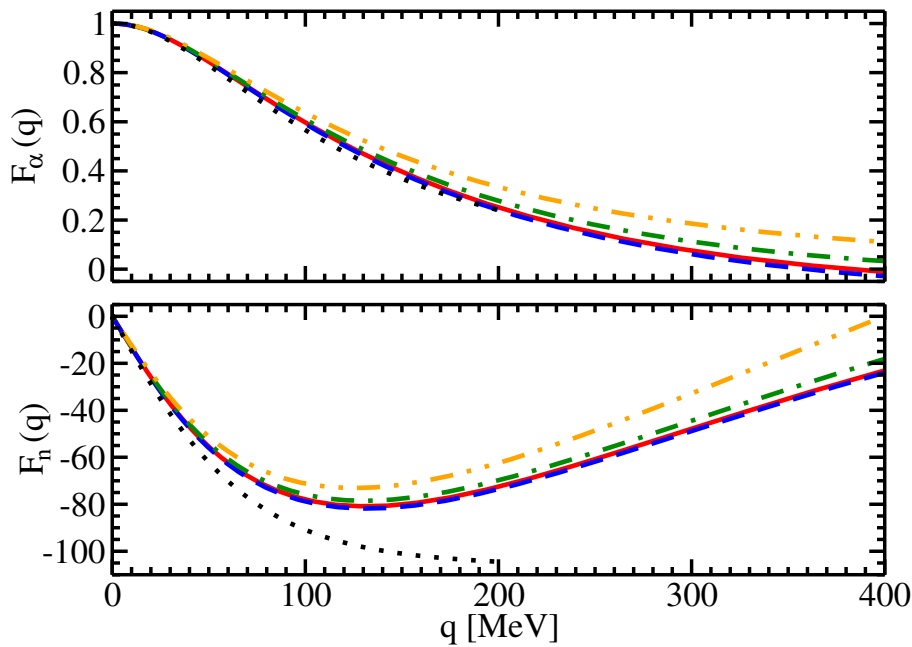}}
\caption{(Color online) The Faddeev components $F_\alpha$ and $F_n$ as functions of $q$, calculated with cutoff parameters $\Lambda$ at 200 MeV (black dotted line), 400 MeV (orange dot-dot-dashed line), 800 MeV (green dot-dashed line), 1.6 GeV (blue dashed line) and 3.2 GeV (red solid line). The Faddeev components are normalized to provide $F_\alpha(0)=1$. }
\label{pic:FcFn}
\end{figure}

After the renormalization, we calculated the Faddeev component $F_n(q)$ from Eq.~\eqref{eq:single-Fn}. By inserting the renormalized $F_n(q)$ into the integrals in Eq.~\eqref{eq:Fc-faddeev}, we can calculate $F_\alpha(q)$ without adding an additional counterterm. Fig.~\ref{pic:FcFn} shows the Faddeev components $F_\alpha$ and $F_n$ as functions of the momentum $q$ for different values of $\Lambda$. The cutoff dependence of the low-$q$ part of both $F_\alpha(q)$ and $F_n(q)$ is weak for $\Lambda > 200$ MeV.

The integral equation (\ref{eq:single-Fn}), modified according to Eq.~(\ref{eq:Xnn-3bf}), is  now renormalized. Moreover, it generates a shallow bound state, with characteristic momenta $\sim M_{lo}$. This seems to contradict the power-counting arguments at the end of the previous subsection. It could be, though, that the binding arises mainly because of short-distance ($\sim 1/M_{hi}$) physics in this EFT, i.e. the ``long-range" ($\sim 1/M_{lo}$) effects of $X_{nn}$ and $I_{n \alpha n}$ are perturbative corrections to a fine-tuned ${}^5$He-$n$ bound state. Whether or not that is the case warrants further investigation. The calculation we have performed here, which only looks at one observable, $B_3$, cannot definitively decide the issue. We are presently examining the correlations among different observables, such as the charge and matter radii of ${}^6$He, calculated with the Faddeev components shown in Fig.~\ref{pic:FcFn}~\cite{Ji:2012he}. The extent to which $X_{nn}$ and $I_{n \alpha n}$ drive those 
correlations will help establish whether the power counting of Ref.~\cite{Bedaque:2003wa} applies in this system. 

Here we have shown the cutoff independence of Faddeev components after renormalization (see Fig.~\ref{pic:FcFn}). This indicates that one three-body parameter (e.g. $B_3$) is needed for renormalization of the LO equations that describe $^6$He in this EFT. In fact, alternative renormalization approaches, e.g., by adding a different three-body counterterm, may be possible. However, any alternative renormalization method must be equivalent to the method used above up to higher-order corrections. The number of three-body renormalization parameters needed for renormalization at LO should not change.

\section{Summary and Outlook}
\label{sec:concl-he6}

We describe the $^6$He ground state as a $nn\alpha$ three-body system in the framework of Halo EFT. The two-body, i.e. $nn$ and $n\alpha$, interactions are expanded under an EFT power counting that produces---at leading order (LO)---the narrow p-wave resonance in the $n\alpha$ ${}^2$P$_{3/2}$ channel and the virtual state in the ${}^1$S$_0$ channel of $nn$ scattering. 
These $nn$ and $n\alpha$ t-matrices are implemented in our LO analysis of  the $^6$He
ground state, which employs a Faddeev formulation to calculate the $^6$He two-neutron
separation energy, $B_3$, as well as the Faddeev components, $F_\alpha$ and $F_n$, via
two coupled integral equations. The result for the ${}^6$He 
two-neutron separation energy
is strongly cutoff dependent. To remedy this we introduce a p-wave
$nn\alpha$ counterterm and perform a renormalization in the three-body $0^+$ sector of
the theory. By tuning the parameter $H_0$ of the three-body counterterm, we can reproduce the
experimental value of  $B_3=0.975$ MeV. The bound-state Faddeev components $F_\alpha$
and $F_n$ are then predicted, and they are both cutoff independent.

The parameter $H_0$ is studied as a function of the cutoff. It exhibits a log-periodic behavior with decreasing periods at large cutoffs. A similar log-periodicity of the three-body-counterterm parameter, however with a constant period, has been observed in the leading-order calculation of three-body systems with s-wave interactions. The different large-$\Lambda$ behavior could be caused by properties of the p-wave $n \alpha$ interactions in the $^6$He system, which breaks the scale-invariance symmetry that is present at LO in the three-body system with only s-wave interactions.

The Halo EFT for ${}^6$He presented here bears a significant similarity to cluster
models. For example, in
Refs.~\cite{Ghovanlou:1974zza,Lehman:1982zz} Ghovanlou and Lehman used separable
Yamaguchi potentials for the $nn$ and $n\alpha$ interactions. They fitted the
parameters to known phase-shifts in these systems and then predicted the binding energy
of ${}^6$He, ultimately obtaining a value smaller than that seen in experiment. In
fact, two-body phase-shifts are insufficient to determine the three-body binding energy
in these systems. This fact  is reflected in the EFT calculation by the sensitivity to the cutoff parameter. The EFT then mandates the introduction of a three-body parameter at LO, and ultimately this may be a  more effective path to a cluster description of the ${}^6$He nucleus than one based on two-body potentials alone.

Of course, there are higher-order corrections in the EFT, which will perturb the result
obtained here. These include the effective-range terms in the $nn$-${}^1$S$_0$ channel
and higher-order effects in the $n\alpha$-$^2$P$_{3/2}$ channel, the role of the
$n\alpha$-${}^2$S$_{1/2}$ and ${}^2$P$_{1/2}$ channels, etc. 
These will be investigated in future work. 
 These higher-order terms can be studied perturbatively using methods similar to those
of Refs.~\cite{Hammer:2001gh,Ji:2011qg,Ji:2012nj,Vanasse:2013sda}. The power counting
of Ref.~\cite{Bedaque:2003wa} indicates that the expansion parameter of the EFT is
$M_{lo}/M_{hi}\sim 1/4$, which is similar to the one for the ``pionless" EFT that has
been applied with much success to few-nucleon systems. However, success there was achieved only after higher-order corrections were included in the analysis, To compare with experimental measurements in these systems at a high accuracy, higher-order effects must be included in the EFT calculation. 

In the renormalization of the $n\alpha$ interaction, we adopt the power counting by
Bedaque {\it et al.}~\cite{Bedaque:2003wa}, i.e. $a_1\sim M_{lo}^{-2}M_{hi}^{-1}$ and
$r_1\sim M_{hi}$, to extract the corresponding p-wave resonance. An alternative power
counting is introduced by Bertulani {\it et al.}~\cite{Bertulani:2002sz} for studying
the $n\alpha$ p-wave interaction. In that work $a_1\sim M_{lo}^{-3}$ and $r_1\sim
M_{lo}$. Therefore, both $\gamma_1$ and $k_{\pm}$ are of order $M_{lo}$, which means both a shallow bound state and a low-energy resonance are present. This result to some extent contradicts the experimentally known absence of the $^5$He bound state. Meanwhile, in contrast to the power counting in Ref.~\cite{Bedaque:2003wa}, which requires one fine-tuning in the renormalization, the power counting in~\cite{Bertulani:2002sz} requires two fine-tunings, which is less ``natural''. It was this power counting that was used by Rotureau and van Kolck in their calculations of $^6$He ground state~\cite{Rotureau:2012yu}.

In that study the authors solved for the Helium-6 bound state using the Gamow shell-model basis. As we did, they introduced a p-wave $nn\alpha$ counterterm in order to render their results independent of the ultraviolet cutoff on the shell-model basis. However, the corresponding three-body parameter vanishes in the limit $\Lambda\rightarrow \infty$, and does not appear to oscillate as a function of $\Lambda$. It thus behaves differently from our three-body parameter. 
Presumably this is a result of the different power counting used for the $n\alpha$ interaction, which leads to different ultraviolet behavior of the integral-equation kernel. 
This deserves further investigation.
 
Our work and Ref.~\cite{Rotureau:2012yu}  both reproduce the experimental value of
$^6$He two-neutron separation energy. A comparison of the
different power counting schemes for ${}^6$He can be achieved if other physical
observables in the $^6$He system can be predicted. In our approach, the Faddeev
components, $F_\alpha$ and $F_n$, are calculated after the renormalization, and are
therefore a prediction. Using $F_\alpha$ and $F_n$, we will be able to construct the
$^6$He wave function and from it the matter-density form factors. These form factors
can be used to obtain predictions for the mean square radii (cf.
Ref.~\cite{Wang:2004ze}) and other observables. The calculation of these quantities is in progress~\cite{Ji:2012he}. The accuracy of the LO Halo-EFT description can also be assessed by examining properties of the $nn \alpha$ system in the continuum. The low-energy $nn \alpha$ continuum was investigated in, e.g. Refs.~\cite{Khaldi:2010jg} within a cluster description, 
as well as in  Refs.~\cite{Pieper:2004qw,Romero-Redondo:2013wma}  via {\it ab initio}
calculations. The Faddeev formalism developed here can be readily extended to continuum
states and used to compute, e.g. LO Halo-EFT predictions for the resonance-pole
positions of excited states in the ${}^6$He system.

In our calculations, the neutron-core mass ratio is kept as a variable ($A$). This
opens up a possible extensions of the current $^6$He analysis to other p-wave halo
nuclei with a different neutron-core mass ratio. One important example is the $^{11}$Li
nucleus, which is another Borromean system but with a $\frac{3}{2}^{-}$ ground state. A recent measurement of the two-neutron transfer reaction, $^1$H($^{11}$Li,$^9$Li)$^3$H, at the ISAC-II facility at TRIUMF, implies that both the s- and p-wave components of $n-^{9}$Li interactions contribute significantly to the ground-state $^{11}$Li~\cite{Tanihata:2008vw}. This suggests that a LO EFT analysis of the $^{11}$Li nucleus should include both the s- and p-wave $n-^{9}$Li interactions non-perturbatively (cf.~Ref.~\cite{Canham:2008jd}), yielding a Faddeev equation that includes more channels than does that for $^6$He at LO. But, similarly to $^6$He, we can also calculate the binding energy and matter radius in $^{11}$Li. In the case of ${}^{11}$Li it will be important to understand whether the presence of additional channels in the LO calculation means that more than one three-body parameter is needed for renormalization at LO.

\acknowledgments{This work was supported by the U.S. Department of Energy under grant DE-FG02-93ER40756 and also in part by both the Natural Sciences and Engineering Research Council (NSERC) and the National Research Council Canada. We thank the Institute for Nuclear Theory at the University of Washington for its hospitality during the program INT-14-1 ``Universality in Few-Body Systems'' when the work was finalized. We are indebted to Lucas Platter, Arbin Thapaliya, Bira van Kolck and Gerry Hale for useful discussions. }
\appendix

\section{Renormalization with Yamaguchi form factors}
\label{app:yamaguchi}
Choosing the form factor $g_l(q)$ 
present in Eq.~\eqref{eq:vl-sep} of Yamaguchi form~\eqref{eq:gl-yamaguchi},
and
assuming $\beta_l\sim M_{hi}$, we can regularize Eq.~\eqref{eq:taul-int} to express
$\tau_l$'s dependence on $M_{hi}$. The two-body coupling constant $\lambda_l$ is then
tuned correspondingly to cancel this divergence. By doing so, we can reproduce the
low-energy behavior of the two-body t-matrices of Eqs.~(\ref{eq:t0-range}, \ref{eq:t1-range}).

For s-wave scattering, $l=0$, we have
\begin{align}
\label{eq:tau0-yama}
\tau_0^{-1}(E) =& \frac{1}{\lambda_0} + 8\pi\mu \int^\infty_0 dq\, \frac{q^2 \beta_0^4}{(q^2-k^2-i\epsilon) (q^2+\beta_0^2)^2} 
\nn
=& \frac{1}{\lambda_0} + 2\pi^2 \mu \frac{\beta_0^3}{(\beta_0-ik)^2}~.
\end{align}
We substitute $\tau_0(E)$ of Eq.~\eqref{eq:tau0-yama} and $g_0(k)$ of
Eq.~\eqref{eq:gl-yamaguchi} into Eq.~\eqref{eq:tl-sep}, and expand the resulting $nn$
scattering t-matrix in powers of $k/\beta_0$ and obtain
\begin{align}
\langle\mathbf{k}|t_{nn}|\mathbf{k'}\rangle =& g_0^{2}(k)\, \tau_0(E)
\nn
=& \left(1+\frac{k^2}{\beta_0^2}\right)^{-2} \left[ \frac{1}{\lambda_0} + 2\pi^2 \mu_{nn} \beta_0\left(1+i\frac{k}{\beta_0}-\frac{k^2}{\beta_0^2}+\cdots\right)^2
\right]^{-1}
\nn
=& \frac{1}{4\pi^2\mu_{nn} } \left\lbrace \left(\frac{\beta_0}{2}+\frac{1}{4\pi^2\mu_{nn} \lambda_0}\right) -\left[\frac{3}{2\beta_0}-\frac{2}{\beta_0^2}\left(\frac{\beta_0}{2}+\frac{1}{4\pi^2\mu_{nn} \lambda_0}\right)\right] k^2 +ik
\right\rbrace^{-1}~. 
\end{align}
By tuning both $\beta_0$ and $\lambda_0$, $a_0$ and $r_0$ are reproduced in the renormalization as 
\begin{subequations}
\begin{align}
\frac{1}{a_0} =&\frac{\beta_0}{2}+\frac{1}{4\pi^2\mu_{nn} \lambda_0}~,
\\
\frac{r_0}{2} =&\frac{3}{2\beta_0}-\frac{2}{\beta_0^2}\,\frac{1}{a_0}~.
\end{align}
\end{subequations}
Note that since $\beta_0\sim M_{hi}$, if $\lambda_0$ is fine tuned in renormalization so that $1/a_0\sim M_{lo}$, then $r_0\sim 1/M_{hi}$ will be naturally obtained.

For the p-wave interaction $l=1$ we calculate $\tau_1$ from Eq.~\eqref{eq:taul-int} as
\begin{align}
\tau_1^{-1}(E) =& \frac{1}{\lambda_1} + 8\pi\mu \int^\infty_0 dq\, \frac{q^4 \beta_1^8}{(q^2-k^2-i\epsilon) (q^2+\beta_1^2)^4} 
\nn
=& \frac{1}{\lambda_1} + \frac{\pi^2 \mu}{4}\beta_1^5\, \frac{\beta_1^2-4i\beta_1 k -k^2}{(\beta_1-ik)^4}~.
\end{align}
We expand $g_1(k)$ and $\tau_1(E)$ in powers of $k$, substitute them into
Eq.~\eqref{eq:tl-sep}, and then obtain the $n\alpha$ scattering t-matrix as
\begin{align}
\langle\mathbf{k}|t_{n\alpha}|\mathbf{k'}\rangle 
=& 3\hat{\mathbf{k}}\cdot\hat{\mathbf{k}}'\, g_1^2(k)\, \tau_1(E) 
\nn
=& 
3\mathbf{k}\cdot\mathbf{k'}
\left(1+\frac{k^2}{\beta_1^2}\right)^{-4}
\left[ \frac{1}{\lambda_1} + \frac{\pi^2 \mu_{n\alpha} }{4}\beta_1^3\,
\left(1-i\frac{4k}{\beta_1}-\frac{k^2}{\beta_1^2}\right)
\left(1+\frac{ik}{\beta_1}-\frac{k^2}{\beta_1^2}+\cdots\right)^4
\right]^{-1}
\nn
=& \frac{3\mathbf{k}\cdot\mathbf{k'}}{4\pi^2\mu_{n\alpha} } 
\left\lbrace \left(\frac{\beta_1^3}{16}+\frac{1}{4\pi^2\mu_{n\alpha} \lambda_1}\right) +\left[\frac{5\beta_1}{16}+\frac{4}{\beta_1^2}\left(\frac{\beta_1^3}{16}+\frac{1}{4\pi^2\mu_{n\alpha} \lambda_1}\right)\right] k^2 +ik^3
\right\rbrace^{-1}~. 
\end{align}
In the p-wave renormalization, $a_1$ and $r_1$ can be reproduced from the relation
\begin{subequations}
\begin{align}
\frac{1}{a_1} =&\frac{\beta_1^3}{16}+\frac{1}{4\pi^2\mu_{n\alpha} \lambda_1}~,
\\
\frac{r_1}{2} =&-\frac{5\beta_1}{16}-\frac{4}{\beta_1^2}\,\frac{1}{a_1}~.
\end{align}
\end{subequations}
Note that since $\beta_1\sim M_{hi}$, if $\lambda_1$ is tuned so that $1/a_1 \sim M_{lo}^2 M_{hi}$, then we can naturally have $r_1\sim M_{hi}$.


\section{The Kernel Functions $X_{n\alpha}$, $X_{\alpha n}$ and $X_{n n}$}
\label{app:kernel}

\subsection{The spin--and--orbital-angular-momentum decomposition}
The spin matrix elements between different spectator representations can be calculated using Wigner's 6-j symbol:
\begin{align}
\left \langle s_1,(s_2 s_3)s_{23}; S\right|\left. (s_1 s_2)s_{12}, s_3; S\right\rangle
=(-1)^{s_1+s_2+s_3+S} \sqrt{(2s_{12}+1)(2s_{23}+1)}
\begin{Bmatrix}
  s_1 & s_2 &  s_{12} \\
  s_3 & S &  s_{23}
\end{Bmatrix}~.
\end{align}
Therefore, in $^6$He's ground state, the spin matrix elements between states represented in either $n$- or $\alpha$-spectator representations are calculated as
\begin{subequations}
\label{eq:spin-matrix}
\begin{align}
&\frac{}{}_{\alpha}\left\langle \left(\frac{1}{2}\, \frac{1}{2}\right)0,0;S_\alpha=0, M_{S\alpha}=0\,\right|
\left.\left(\frac{1}{2}0\right)\frac{1}{2},\frac{1}{2};\,S_n=L_n,M_{Sn}=-M_{Ln}\right\rangle_{n}
=-\delta_{0,L_n} \delta_{0,M_{Ln}}~,
\\
&\frac{}{}_{n}\left\langle\left(\frac{1}{2}0\right)\frac{1}{2},\frac{1}{2};\,S_n=L_n,M_{Sn}=-M_{Ln}\right|
\left.\left(\frac{1}{2}\, \frac{1}{2}\right) \,0,0;S_\alpha=0, M_{S\alpha}=0\,\right\rangle_{\alpha}
=-\delta_{0,L_n} \delta_{0,M_{Ln}}~,
\\
&\frac{}{}_{n}\left\langle\left(\frac{1}{2}0\right)\frac{1}{2},\frac{1}{2};\,S_n=L_n,M_{Sn}=-M_{Ln}\right|\mathcal{P}_{nn}
\left|\left(\frac{1}{2}0\right)\frac{1}{2},\frac{1}{2};\,S_n=L_n',M_{Sn}=-M_{Ln'}\right\rangle_{n}
\nonumber\\
&\hskip 8 cm =(-1)^{1-L_n} \delta_{L_n,L_n'}\, \delta_{M_{Ln},M_{Ln'}}~.
\end{align}
\end{subequations}

By substituting Eqs.~\eqref{eq:spin-matrix} into Eqs. (\ref{eq:state-alpha-LS}, \ref{eq:state-n-LS}), we can decompose the spin and orbital-angular-momentum parts of the matrix elements $\,_{i}\langle p,q;\Omega_i | p',q';\Omega_j \rangle_{j}$ in the $^6$He system as
\begin{subequations}
\label{eq:omega-matrix}
\begin{align}
\,_{\alpha}\langle p,q;\Omega_\alpha | p',q';\Omega_n \rangle_{n}
=& -\sqrt{\frac{2}{3}}\, 
\,_{\alpha}\langle p,q;\,0,0; L_\alpha=M_{L\alpha}=0 |
p',q';\,1,1;\,L_n=M_{Ln}=0\rangle_{n}~,
\\
\,_{n}\langle p,q;\Omega_n | p',q';\Omega_\alpha \rangle_{\alpha}
=& -\sqrt{\frac{2}{3}}\, 
\,_{n}\langle p,q;\,0,0; L_n=M_{Ln}=0 |
p',q';\,1,1;\,L_\alpha=M_{L\alpha}=0\rangle_{\alpha}~,
\\
\,_{n}\langle p,q;\Omega_n |-\mathcal{P}_{nn}| p',q';\Omega_n \rangle_{n}
=& \sum\limits_{L_n=0}^1 \sum\limits_{M_{Ln}=-L_n}^{L_n}\frac{(-2)^{1-L_n}}{6L_n+3}
\,_{n}\langle p,q;\,0,0; L_n, M_{Ln} |-\mathcal{P}_{nn}|
p',q';\,1,1;\,L_n, M_{Ln}\rangle_{n}~. 
\end{align}
\end{subequations}

Inserting Eqs.~\eqref{eq:omega-matrix} into Eq.~\eqref{eq:faddeev3}, we can decouple the Kernel functions, $X_{ij}$, in the $^6$He problem into a summation of functions $\mathcal{Z}_{ij}^{(L)}$ at different overall orbital angular momentum $L$:
\begin{subequations}
\label{eq:X-Z}
\begin{align}
X_{\alpha n}(q,q';E) =& -\sqrt{\frac{2}{3}} \mathcal{Z}_{\alpha n}^{(0)}(q,q';E)~,
\\
X_{n\alpha}(q,q';E) =& -\sqrt{\frac{2}{3}} \mathcal{Z}_{n\alpha}^{(0)}(q,q';E)~,
\\
X_{nn}(q,q';E) =& -\frac{2}{3} \mathcal{Z}_{nn}^{(0)}(q,q';E) +\frac{1}{3}\mathcal{Z}_{nn}^{(1)}(q,q';E)~.
\end{align}
\end{subequations}
In the spectator-$\alpha$ representation $L=0$; while in the spectator-$n$ representation $L$ can be zero or one. Those functions $\mathcal{Z}_{ij}^{(L)}$ are then
\begin{subequations}
\begin{align}
\mathcal{Z}_{\alpha n}^{(0)}(q,q';E) =& 
\iint p^2 dp\, p'^2 d p'\, g_0(p) G_0^{(\alpha)}(p,q;E)\, g_1(p')
 \,_{\alpha}\langle p,q;\,(00)00\, |
\,p',q';\,(11)00\rangle_{n}~,
\\
\mathcal{Z}_{n\alpha}^{(0)}(q,q';E) =&
\iint p^2 dp\, p'^2 d p'\, g_1(p) G_0^{(n)}(p,q;E)\, g_0(p')
 \,_{n}\langle p,q;\,(11)00\, |
\,p',q';\,(00)00\rangle_{\alpha}~,
\\
\mathcal{Z}_{nn}^{(L)}(q,q';E) =&
\iint p^2 dp\, p'^2 d p'\, g_1(p) G_0^{(n)}(p,q;E)\, g_1(p')
\,_{n}\langle p,q;\,(11)LM\, |-\mathcal{P}_{nn}|
\,p',q';\,(11)LM\rangle_{n}~.
\end{align}
\end{subequations}
$\mathcal{Z}_{nn}^{(L)}$ is independent of the quantum number $M$ for both the $L=0$ and $L=1$ cases, which will be proved later.

\subsection{The funcions $\mathcal{Z}_{\alpha n}^{(0)}$, $\mathcal{Z}_{n\alpha }^{(0)}$ and $\mathcal{Z}_{nn }^{(L)}$}

Here we we calculate the orbital-angular-momentum--dependent kernel functions $\mathcal{Z}_{\alpha n}^{(0)}$, $\mathcal{Z}_{n\alpha }^{(0)}$ and $\mathcal{Z}_{nn }^{(L)}$.

\subsubsection{The Function $\mathcal{Z}_{\alpha n}^{(0)}$}
After inserting two complete sets of Jacobi-momentum states, we can write $\mathcal{Z}_{\alpha n}^{(0)}$ as
\begin{align}
\label{eq:Z-cn-0}
\mathcal{Z}_{\alpha n}^{(0)}(q,q';E) =& 
\iint p^2 dp\, p'^2 d p'\, g_0(p) G_0^{(\alpha)}(p,q;E)\, g_1(p')
\iint d^3 p_1 d^3q_1 \iint d^3 p_2 d^3q_2
\nn
&\times \,_{\alpha}\langle p,q;\,(00)00\, |\mathbf{p_1}\mathbf{q_1}\rangle_\alpha \,_{\alpha}\langle\mathbf{p_1}\mathbf{q_1}|\mathbf{p_2}\mathbf{q_2}\rangle_{n} \,_{n}\langle \mathbf{p_2}\mathbf{q_2}|
\,p',q';\,(11)00\rangle_{n}~,
\end{align}
where the matrix elements containing the orbital-angular-momentum quantum numbers can be expressed as
\begin{subequations}
\label{eq:Yc-Yn}
\begin{align}
\,_{\alpha}\langle \mathbf{p_1}\mathbf{q_1}|p,q;\,(00)00\rangle_\alpha =& \frac{1}{p_1^2}\delta(p_1-p) \frac{1}{q_1^2}\delta(q_1-q) \mathcal{Y}_{00}^{00}(\hat{\mathbf{p}}_1 \hat{\mathbf{q}}_1)~,
\\
\,_{n}\langle \mathbf{p_2}\mathbf{q_2}|p',q';\,(11)00\rangle_{n} =&
\frac{1}{p_2^2}\delta(p_2-p') \frac{1}{q_2^2}\delta(q_2-q') \mathcal{Y}_{11}^{00}(\hat{\mathbf{p}}_2 \hat{\mathbf{q}}_2)~.
\end{align}
\end{subequations}
The function $\mathcal{Y}_{l_1 l_2}^{LM}$ indicates the orbital-angular-momentum coupling of two spherical harmonics to produce an overall orbital angular momentum $L$ and z-component $M$:
\begin{equation}
\mathcal{Y}_{l_1 l_2}^{LM}(\hat{\mathbf{q}}_1\hat{\mathbf{q}}_2) =\sum\limits_{m_1 m_2} C(l_1 l_2 L|m_1 m_2 M)\, \textrm{Y}_{l_1 m_1}(\hat{\mathbf{q}}_1) \,\textrm{Y}_{l_2 m_2}(\hat{\mathbf{q}}_2)~.
\end{equation}

Also, the transition between the free momentum states $|\mathbf{p_1}\mathbf{q_1}\rangle_{\alpha}$ and $|\mathbf{p_2}\mathbf{q_2}\rangle_{n}$ yields the product of two delta functions:
\begin{equation}
\,_{\alpha}\langle\mathbf{p_1}\mathbf{q_1}|\mathbf{p_2}\mathbf{q_2}\rangle_{n}
=\delta^{(3)}(\mathbf{p}_1-\mathbf{P}_{\alpha n}) \delta^{(3)}(\mathbf{p}_2+\mathbf{P'}_{\alpha n})~,
\end{equation}
with
\begin{subequations}
\label{eq:delta-cn}
\begin{align}
\mathbf{P}_{\alpha n} =& \frac{\mu_{nn}}{m_n}\mathbf{q}_1 + \mathbf{q}_2 = \frac{1}{2}\mathbf{q}_1 + \mathbf{q}_2~,
\\
\mathbf{P}_{\alpha n}' =& \mathbf{q}_1 + \frac{\mu_{n\alpha}}{m_n}\mathbf{q}_2 = \mathbf{q}_1 + \frac{A}{A+1}\mathbf{q}_2~,
\end{align}
\end{subequations}
where $q_1=q$ and $q_2=q'$ are determined from Eqs.~\eqref{eq:Yc-Yn}.

By applying Eqs.~(\ref{eq:Yc-Yn}--\ref{eq:delta-cn}) into Eq.~\eqref{eq:Z-cn-0}, we obtain
\begin{align}
\label{eq:Z-cn-1}
\mathcal{Z}_{\alpha n}^{(0)}(q,q';E) =& 
\int d\hat{\mathbf{q}}_1 \int d\hat{\mathbf{q}}_2\,
g_0(P_{\alpha n}) G_0^{(\alpha)}(P_{\alpha n},q;E)\, g_1(P_{\alpha n}')
\mathcal{Y}_{00}^{00\,*}(\hat{\mathbf{P}}_{\alpha n} \hat{\mathbf{q}}_1)\,
\mathcal{Y}_{11}^{00}(-\hat{\mathbf{P}}_{\alpha n}'\, \hat{\mathbf{q}}_2)
\nn
=& \frac{1}{4\pi} \int d\hat{\mathbf{q}}_1 \int d\hat{\mathbf{q}}_2\,
g_0(P_{\alpha n}) G_0^{(\alpha)}(P_{\alpha n},q;E)\, g_1(P_{\alpha n}') 
\sum\limits_{m=-1}^{1} C(110|m\,-m\,0)
\textrm{Y}_{1m}(-\hat{\mathbf{P}}_{\alpha n}')\, \textrm{Y}_{1-m}(\hat{\mathbf{q}}_2)~,
\end{align}
where we used the fact that $\mathcal{Y}_{00}^{00}(\hat{\mathbf{P}}_{\alpha n} \hat{\mathbf{q}}_1)=1/(4\pi)$.

Using the relation~\cite{Glockle:1983}
\begin{align}
\textrm{Y}_{l m}(\widehat{\mathbf{r_1}+\mathbf{r_2}}) =& \sum\limits_{l_1+l_2=l}\sqrt{\frac{4\pi\,(2l+1)!}{(2l_1+1)!\,(2l_2+1)!}} \,
\frac{r_1^{l_1} r_2^{l_2}}{|\mathbf{r_1}+\mathbf{r_2}|^l}
\sum\limits_{m_1 m_2} C(l_1 l_2 l| m_1 m_2 m)\, \textrm{Y}_{l_1 m_1}(\hat{\mathbf{r}}_1)\,\textrm{Y}_{l_2 m_2}(\hat{\mathbf{r}}_2)~,
\end{align}
we rewrite Eq.~\eqref{eq:Z-cn-1} as
\begin{align}
\label{eq:Z-cn-2}
\mathcal{Z}_{\alpha n}^{(0)}(q,q';E) =& 
\frac{1}{4\pi}
\int d\hat{\mathbf{q}}_1 \int d\hat{\mathbf{q}}_2\,
g_0(P_{\alpha n}) G_0^{(\alpha)}(P_{\alpha n},q;E)\, g_1(P_{\alpha n}') \,P_{\alpha n}'^{-1}
\nn
&\times 
\sum\limits_{m=-1}^{1} C(110|m\,-m\,0)
\sum\limits_{l_1+l_2=1}\sqrt{\frac{4\pi\,3!}{(2l_1+1)!\,(2l_2+1)!}} \,q^{l_1} \left(\frac{A}{A+1}q'\right)^{l_2}
\nn
&\times
\sum\limits_{m_1 m_2} C(l_1 l_2 1| m_1 m_2 m)
\textrm{Y}_{l_1 m_1}(-\hat{\mathbf{q}}_1)\,\textrm{Y}_{l_2 m_2}(-\hat{\mathbf{q}}_2)
\textrm{Y}_{1-m}(\hat{\mathbf{q}}_2)~.
\end{align}

We can perform a Legendre expansion of the product of terms in front of the first summation in Eq.~\eqref{eq:Z-cn-2}. Since the Halo EFT calculation takes $g_0(P_{\alpha n})=1$ and $g_1(P_{\alpha n}')=P_{\alpha n}'$, we have
\begin{align}
g_0(P_{\alpha n}) G_0^{(\alpha)}(P_{\alpha n},q;E)\, g_1(P_{\alpha n}') \,P_{\alpha n}'^{-1}=G_0^{(\alpha)}(P_{\alpha n},q;E)
=4\pi \sum\limits_{t \nu}\mathcal{G}_{\alpha n}^{\,t}(q,q';E) \,\textrm{Y}_{t\nu}^{\,*}(\hat{\mathbf{q}}_1)
\textrm{Y}_{t\nu}(\hat{\mathbf{q}}_2)~,
\end{align}
where $\mathcal{G}_{\alpha n}^{\,t}$ is determined by
\begin{align}
\label{eq:G-cn-1}
\mathcal{G}_{\alpha n}^{\,t}(q,q';E) =& \frac{1}{2}\int^{1}_{-1} dx\, \textrm{P}_t(x)\, G_0^{(\alpha)}(P_{\alpha n},q;E)
\nn
=&\frac{1}{2}\int^{1}_{-1} dx\, \textrm{P}_t(x)\left[ E-\frac{1}{m_n}\left(\frac{1}{4}q^2+q'^2 + q q' x\right) -\frac{A+2}{4A\, m_n}q^2\right]^{-1}
\nn
=&\frac{m_n}{2qq'}\int^{1}_{-1} dx\, \textrm{P}_t(x)\left[\frac{1}{qq'}\left(m_n E-\frac{A+1}{2A}q^2-q'^2\right) -x\right]^{-1}~.
\end{align}

If we define
\begin{equation}
\label{eq:G-cn-2}
z_{\alpha n} = \frac{1}{qq'}\left(m_n E-\frac{A+1}{2A}q^2-q'^2\right)~,
\end{equation}
we can relate $\mathcal{G}_{\alpha n}^{\,t}(q,q';E)$ to the Legendre functions of the second kind $\textrm{Q}_t$ [see Eq.~\eqref{eq:Ql}] as
\begin{equation}
\label{eq:G-cn-3}
\mathcal{G}_{\alpha n}^{\,t}(q,q';E) = \frac{m_n}{qq'}\,\textrm{Q}_t (z_{\alpha n})~.
\end{equation}
Here we take the opportunity to write explicitly the three $\textrm{Q}_l$'s used in our calculation as
\begin{subequations}
\begin{align}
\textrm{Q}_0(z) =& \frac{1}{2} \ln \left(\frac{z+1}{z-1}\right)
\\
\textrm{Q}_1(z) =& \frac{1}{2}z \ln \left(\frac{z+1}{z-1}\right) -1
\\
\textrm{Q}_2(z) =& \frac{1}{2}\left(-\frac{1}{2}+\frac{3}{2}z^2\right) \ln \left(\frac{z+1}{z-1}\right) -\frac{3}{2}z ~.
\end{align}
\end{subequations}

Now the dependences on $\hat{\mathbf{q}_1}$ and $\hat{\mathbf{q}_2}$ are separated
and can be integrated individually as
\begin{align}
\mathcal{Z}_{\alpha n}^{(0)}(q,q';E) =& \sum\limits_{t}
\mathcal{G}_{\alpha n}^{\,t}(q,q';E)
\sum\limits_{m=-1}^{1} C(110|m\,-m\,0)
\sum\limits_{l_1+l_2=1}\sqrt{\frac{4\pi\,3!}{(2l_1+1)!\,(2l_2+1)!}}
\nn
&\times 
 q^{l_1} \left(\frac{A}{A+1}q'\right)^{l_2}
\sum\limits_{m_1 m_2} C(l_1 l_2 1| m_1 m_2 m)
\nn
&\times  \sum\limits_{\nu=-t}^{t}
\int d\hat{\mathbf{q}}_1\, 
\textrm{Y}_{t\nu}^{\,*}(\hat{\mathbf{q}}_1)
\textrm{Y}_{l_1 m_1}(-\hat{\mathbf{q}}_1)
\int d\hat{\mathbf{q}}_2\,
\textrm{Y}_{t\nu}(\hat{\mathbf{q}}_2)
\textrm{Y}_{l_2 m_2}(-\hat{\mathbf{q}}_2)
\textrm{Y}_{1-m}(\hat{\mathbf{q}}_2)~.
\end{align}

After integrating the product of spherical harmonics, we sum up all the orbital-angular-momentum quantum numbers using properties of Clebsch-Gordan coefficients (see e.g., in Ref.~\cite{Edmonds:1957}), 
and express $\mathcal{Z}_{\alpha n}^{(0)}$ as a summation of $\mathcal{G}_{\alpha n}^{\,t}$'s:
\begin{equation}
\label{eq:Z-cn}
\mathcal{Z}_{\alpha n}^{(0)}(q,q';E)
= \sqrt{3}\left(\frac{A}{A+1} q'\, \mathcal{G}_{\alpha n}^{\,0}(q,q';E) +q\, \mathcal{G}_{\alpha n}^{\,1}(q,q';E)\right)~.
\end{equation}

\subsubsection{The Function $\mathcal{Z}_{n\alpha }^{(0)}$}
Similarly, $\mathcal{Z}_{n\alpha }^{(0)}$ is calculated as
\begin{align}
\label{eq:Z-nc-0}
\mathcal{Z}_{n\alpha}^{(0)}(q,q';E) =& 
\iint p^2 dp\, p'^2 d p'\, g_1(p) G_0^{(n)}(p,q;E)\, g_0(p')
\iint d^3 p_1 d^3 q_1 \iint d^3 p_2 d^3 q_2
\nn
&\times \,_{n}\langle p,q;\,(11)00\, |\mathbf{p_1}\mathbf{q_1}\rangle_n \,_{n}\langle\mathbf{p_1}\mathbf{q_1}|\mathbf{p_2}\mathbf{q_2}\rangle_{\alpha}
\,_{\alpha}\langle \mathbf{p_2}\mathbf{q_2}|
\,p',q';\,(00)00\rangle_{\alpha}~.
\end{align}

We express the orbital-angular-momentum dependent matrix elements as
\begin{subequations}
\label{eq:Yn-Yc}
\begin{align}
\,_{n}\langle \mathbf{p_1}\mathbf{q_1}|p,q;\,(11)00\rangle_n =& \frac{1}{p_1^2}\delta(p_1-p) \frac{1}{q_1^2}\delta(q_1-q) \mathcal{Y}_{11}^{00}(\hat{\mathbf{p}}_1 \hat{\mathbf{q}}_1)~,
\\
\,_{\alpha}\langle \mathbf{p_2}\mathbf{q_2}|p',q';\,(00)00\rangle_{\alpha} =&
\frac{1}{p_2^2}\delta(p_2-p') \frac{1}{q_2^2}\delta(q_2-q') \mathcal{Y}_{00}^{00}(\hat{\mathbf{p}}_2 \hat{\mathbf{q}}_2)~.
\end{align}
\end{subequations}
Also the transition between momentum states $|\mathbf{p_1}\mathbf{q_1}\rangle_{n}$ and $|\mathbf{p_2}\mathbf{q_2}\rangle_{\alpha}$ yields
\begin{equation}
\,_{n}\langle\mathbf{p_1}\mathbf{q_1}|\mathbf{p_2}\mathbf{q_2}\rangle_{\alpha}
=\delta^{(3)}(\mathbf{p}_1+\mathbf{P}_{n\alpha}) \delta^{(3)}(\mathbf{p}_2-\mathbf{P'}_{n\alpha})~,
\end{equation}
where
\begin{subequations} 
\begin{align}
\mathbf{P}_{n\alpha} =& \frac{\mu_{n\alpha}}{m_n}\mathbf{q}_1 + \mathbf{q}_2 = \frac{A}{A+1}\mathbf{q}_1 + \mathbf{q}_2~,
\\
\mathbf{P}_{n\alpha}' =& \mathbf{q}_1 + \frac{\mu_{nn}}{m_n}\mathbf{q}_2 = \mathbf{q}_1 + \frac{1}{2}\mathbf{q}_2~,
\end{align}
\end{subequations}
with $q_1=q$ and $q_2=q'$ determined from Eqs.~\eqref{eq:Yn-Yc}.

Therefore, we can rewrite Eq.~\eqref{eq:Z-nc-0} as
\begin{align}
\label{eq:Z-nc-1}
\mathcal{Z}_{n\alpha}^{(0)}(q,q';E) =& 
\int d\hat{\mathbf{q}}_1 \int d\hat{\mathbf{q}}_2\,
g_1(P_{n\alpha}) G_0^{(n)}(P_{n\alpha},q;E)\, g_0(P_{n\alpha}')
\mathcal{Y}_{11}^{00\,*}(-\hat{\mathbf{P}}_{n\alpha} \hat{\mathbf{q}}_1)\,
\mathcal{Y}_{00}^{00}(\hat{\mathbf{P}}_{n\alpha}'\, \hat{\mathbf{q}}_2)~,
\end{align}
where 
$\mathcal{Y}_{00}^{00}(\hat{\mathbf{P}}_{n\alpha}'\, \hat{\mathbf{q}}_2)=1/(4\pi)$.

Similarly to Eq.~(\ref{eq:G-cn-1} -- \ref{eq:G-cn-3}), we define a function $\mathcal{G}_{n\alpha}^{t}$ which satisfies
\begin{align}
\mathcal{G}_{n\alpha}^{t} =&\frac{1}{2}\int_{-1}^{1} dx\,\textrm{P}_t(x)\, P_{n\alpha}^{-1}g_1(P_{n\alpha}) G_0^{(n)}(P_{n\alpha},q;E)\, g_0(P_{\alpha n}')
\nn
=&\frac{1}{2}\int_{-1}^{1} dx\,\textrm{P}_t(x)\, G_0^{(n)}(P_{n\alpha},q;E)
\nn
=&\frac{1}{2}\int_{-1}^{1} dx\,\textrm{P}_t(x)
\left[E - \frac{A+1}{2Am_n}\left(\frac{A^2}{(A+1)^2} q^2 +q'^2 + \frac{2A}{A+1} qq'x\right) -\frac{A+2}{2(A+1)m_n}q^2\right]^{-1}
\nn
=&\frac{m_n}{qq'}\, \textrm{Q}_t(z_{n\alpha})
\end{align}
with
\begin{equation}
z_{n\alpha} = \frac{1}{qq'}\left(m_n E - q^2 -\frac{A+1}{2A} q'^2\right)~.
\end{equation}

After similar procedures to those used in calculating $\mathcal{Z}_{\alpha n}^{(0)}$, we can express $\mathcal{Z}_{n \alpha}^{(0)}$ as a summation of $\mathcal{G}_{n \alpha}^{\,t}$'s by
\begin{equation}
\label{eq:Z-nc}
\mathcal{Z}_{n\alpha}^{(0)}(q,q';E)
= \sqrt{3}\left(\frac{A}{A+1} q\, \mathcal{G}_{n\alpha}^{\,0}(q,q';E) +q'\, \mathcal{G}_{n\alpha}^{\,1}(q,q';E)\right)~.
\end{equation}

\subsubsection{The Function $\mathcal{Z}_{nn}^{(L)}$ with $L=0,1$}
Also similarly, $\mathcal{Z}_{nn}^{(L)}$ is calculated as
\begin{align}
\label{eq:Z-nn-0}
\mathcal{Z}_{nn}^{(L)}(q,q';E) &=
\iint p^2 dp\, p'^2 d p'\, g_1(p) G_0^{(n)}(p,q;E)\, g_1(p')
\iint d^3 p_1 d^3 q_1 \iint d^3 p_2 d^3 q_2
\nn
&\times \,_{n}\langle p,q;\,(11)LM\, |\mathbf{p_1}\mathbf{q_1}\rangle_n 
\,_{n}\langle\mathbf{p_1}\mathbf{q_1}|-\mathcal{P}_{nn}|\mathbf{p_2}\mathbf{q_2}\rangle_{n}
\,_{n}\langle \mathbf{p_2}\mathbf{q_2}|\,p',q';\,(11)LM\rangle_{n}~.
\end{align}

The orbital-angular-momentum dependent matrix elements are written as
\begin{align}
\label{eq:Yn-Yn}
\,_{n}\langle \mathbf{p_1}\mathbf{q_1}|p,q;\,(11)LM\rangle_n =& \frac{1}{p_1^2}\delta(p_1-p) \frac{1}{q_1^2}\delta(q_1-q) \mathcal{Y}_{11}^{LM}(\hat{\mathbf{p}}_1 \hat{\mathbf{q}}_1)~.
\end{align}
The transition of momentum states in this case leads to
\begin{equation}
\,_{n}\langle\mathbf{p_1}\mathbf{q_1}|-\mathcal{P}_{nn}|\mathbf{p_2}\mathbf{q_2}\rangle_{n}
=-\delta^{(3)}(\mathbf{p}_1-\mathbf{P}_{nn}) \delta^{(3)}(\mathbf{p}_2-\mathbf{P'}_{nn})~,
\end{equation}
where 
\begin{subequations}
\begin{align}
\mathbf{P}_{nn} =& \frac{\mu_{n\alpha}}{m_\alpha}\mathbf{q}_1 + \mathbf{q}_2 = \frac{1}{A+1}\mathbf{q}_1 + \mathbf{q}_2~,
\\
\mathbf{P}_{nn}' =& \mathbf{q}_1 + \frac{\mu_{n\alpha}}{m_\alpha}\mathbf{q}_2 = \mathbf{q}_1 + \frac{1}{A+1}\mathbf{q}_2~,
\end{align}
\end{subequations}
with $q_1=q$ and $q_2=q'$ determined from Eq.~\eqref{eq:Yn-Yn}.

We then rewrite Eq.~\eqref{eq:Z-nn-0} as
\begin{align}
\label{eq:Z-nn-1}
\mathcal{Z}_{nn}^{(L)}(q,q';E) =& 
-\int d\hat{\mathbf{q}}_1 \int d\hat{\mathbf{q}}_2\,
g_1(P_{nn}) G_0^{(n)}(P_{nn},q;E)\, g_1(P_{nn}')
\nn
&\times
\mathcal{Y}_{11}^{LM\,*}(\hat{\mathbf{P}}_{nn} \hat{\mathbf{q}}_1)\,
\mathcal{Y}_{11}^{LM}(\hat{\mathbf{P}}_{nn}'\, \hat{\mathbf{q}}_2)~.
\end{align}

As in Eq.~(\ref{eq:G-cn-1} -- \ref{eq:G-cn-3}), we define the function $\mathcal{G}_{nn}^{t}$ that satisfies:
\begin{align}
\mathcal{G}_{nn}^{t} =&\frac{1}{2}\int_{-1}^{1} dx\,\textrm{P}_t(x)\, P_{nn}^{-1}g_1(P_{nn}) G_0^{(n)}(P_{nn},q;E)\, g_1(P_{\alpha n}')P_{nn}'^{-1}
\nn
=&\frac{1}{2}\int_{-1}^{1} dx\,\textrm{P}_t(x)\, G_0^{(n)}(P_{nn},q;E)
\nn
=&\frac{1}{2}\int_{-1}^{1} dx\,\textrm{P}_t(x)
\left[E - \frac{A+1}{2Am_n}\left(\frac{q^2}{(A+1)^2}  +q'^2 + \frac{2qq'x}{A+1} \right) -\frac{A+2}{2(A+1)m_n}q^2\right]^{-1}
\nn
=&\frac{m_n}{qq'}\, \textrm{Q}_t(z_{nn})~,
\end{align}
with
\begin{equation}
z_{nn} = \frac{A}{qq'}\left[m_n E - \frac{A+1}{2A}(q^2+q'^2)\right]~.
\end{equation}

Applying similar procedures again we express $\mathcal{Z}_{n n}^{(L)}$ as a summation of $\mathcal{G}_{n n}^{\,t}$'s. For $L=0$ and $L=1$, we obtain
\begin{subequations}
\label{eq:Z-nn}
\begin{align}
\mathcal{Z}_{nn}^{(0)}(q,q';E) =&-3\left[\frac{A^2+2A+4}{3(A+1)^2}qq'\,\mathcal{G}_{nn}^{0}(q,q';E)
+\frac{1}{A+1}(q^2+q'^2)\,\mathcal{G}_{nn}^{1}(q,q';E)
+\frac{2}{3}qq'\,\mathcal{G}_{nn}^{2}(q,q';E)\right]
\\
\mathcal{Z}_{nn}^{(1)}(q,q';E) =& qq'\,\mathcal{G}_{nn}^{0}(q,q';E)-qq'\,\mathcal{G}_{nn}^{2}(q,q';E)~.
\end{align}
\end{subequations}

By substituting Eqs.~(\ref{eq:Z-cn}), (\ref{eq:Z-nc}), and (\ref{eq:Z-nn}) into Eq.~\eqref{eq:X-Z}, we obtain the expressions for the kernel functions $X_{\alpha n}$, $X_{n\alpha}$ and $X_{nn}$ given in Eqs.~\eqref{eq:faddeev-X}.




\begin{thebibliography}{300}

\expandafter\ifx\csname natexlab\endcsname\relax\def\natexlab#1{#1}\fi
\expandafter\ifx\csname bibnamefont\endcsname\relax
  \def\bibnamefont#1{#1}\fi
\expandafter\ifx\csname bibfnamefont\endcsname\relax
  \def\bibfnamefont#1{#1}\fi
\expandafter\ifx\csname citenamefont\endcsname\relax
  \def\citenamefont#1{#1}\fi
\expandafter\ifx\csname url\endcsname\relax
  \def\url#1{\texttt{#1}}\fi
\expandafter\ifx\csname urlprefix\endcsname\relax\def\urlprefix{URL }\fi
\providecommand{\bibinfo}[2]{#2}
\providecommand{\eprint}[2][]{\url{#2}}

\bibitem{Tanihata:1995yv} 
  I.~Tanihata,
  J.\ Phys.\ G {\bf 22}, 157 (1996), and references therein.

\bibitem{Jensen:2004} 
  A.~S.~Jensen, K.~Riisager, D.~V.~Fedorov and E.~Garrido,
  Rev.\ Mod.\ Phys.\ \ {\bf 76}, 215  (2004), and references therein.
  
  
\bibitem{Zhukov:1993aw} 
  M.~V.~Zhukov, B.~V.~Danilin, D.~V.~Fedorov, J.~M.~Bang, I.~J.~Thompson, and J.~S.~Vaagen,
  Phys.\ Rept.\  {\bf 231}, 151 (1993).


\bibitem{Mazumdar:2000dg} 
  I.~Mazumdar, V.~Arora and V.~S.~Bhasin,
  Phys.\ Rev.\ C {\bf 61}, 051303 (2000).
  
\bibitem{Hagen:2013jqa} 
  G.~Hagen, P.~Hagen, H.~-W.~Hammer and L.~Platter,
  Phys.\ Rev.\ Lett.\  {\bf 111}, 132501 (2013).
 
  \bibitem{Pieper:2004qh}
  S.~C.~Pieper,
  Nucl.\ Phys.\ A {\bf 751}, 516 (2005).
  
  \bibitem{Quaglioni:2013kma} 
  S.~Quaglioni, C.~Romero-Redondo and P.~Navr\'{a}til,
  Phys.\ Rev.\ C {\bf 88}, 034320 (2013).
  
  \bibitem{Bacca:2012up} 
  S.~Bacca, N.~Barnea and A.~Schwenk,
  Phys.\ Rev.\ C {\bf 86}, 034321 (2012).
  
  
\bibitem{Brodeur:2012zz} 
  M.~Brodeur
  {\it et al.},
  Phys.\ Rev.\ Lett.\  {\bf 108}, 052504 (2012).
  

\bibitem{Wang:2004ze}
  L.~-B.~Wang
{\it et al.},
  Phys.\ Rev.\ Lett.\  {\bf 93}, 142501 (2004).
  
  
\bibitem{Tanihata:1992wf} 
  I.~Tanihata, D.~Hirata, T.~Kobayashi, S.~Shimoura, K.~Sugimoto and H.~Toki,
  Phys.\ Lett.\ B {\bf 289}, 261 (1992).
  
\bibitem{Alkhazov:1997zz} 
  G.~D.~Alkhazov
  {\it et al.},
  Phys.\ Rev.\ Lett.\  {\bf 78}, 2313 (1997).

\bibitem{Kiselev:2005}
O.~A.~Kiselev
{\it et al.}, 
Eur.\ Phys.\ J.\ A {\bf25}, s01, 215 (2005).


\bibitem{Ghovanlou:1974zza} 
  A.~Ghovanlou, and D.~R.~Lehman,
  Phys.\ Rev.\ C {\bf 9}, 1730 (1974).

\bibitem{Lehman:1982zz}
  D.~R.~Lehman,
  Phys.\ Rev.\ C {\bf 25}, 3146 (1982).


  
  \bibitem{Funada:1994}
    S.~Funada, H.~Kameyama and Y.~Sakuragi,
  Nucl.\ Phys.\ A {\bf 575}, 93 (1994).

\bibitem{Varga:1994}
  K.~Varga, Y.~Suzuki and Y.~Ohbayasi,
  Phys.\ Rev.\ C {\bf 50}, 189 (1994).



\bibitem{Bedaque:1999ve} 
  P.~F.~Bedaque, H.~W.~Hammer, and U.~van Kolck,
  Nucl.\ Phys.\ A {\bf 676}, 357 (2000).

  
\bibitem{Bedaque:1998kg}
P.~F.~Bedaque, H.-W.~Hammer, and U.~van Kolck,
        Phys.\ Rev.\ Lett.\  {\bf 82}, 463 (1999);
        Nucl.\ Phys.\ A {\bf 646}, 444 (1999).





\bibitem{Braaten:2004rn}
  E.~Braaten, and H.~W.~Hammer,
  Phys.\ Rept.\  {\bf 428}, 259 (2006).


\bibitem{Efimov70}
V.~Efimov,
\newblock { Phys. Lett.}, {\bf 33B}, 563 (1970).



\bibitem{Gross:2009}
N.~Gross, Z.~Shotan, S.~Kokkelmans, and L.~Khaykovich,
Phys.\ Rev.\ Lett.\ {\bf 103}, 163202 (2009).


\bibitem{Pollack:2009}
S.~E. Pollack, D. Dries, and R.~G. Hulet,
\newblock {Science}, {\bf 326}, 1683 (2009).


\bibitem{Kraemer:2006}
T. Kraemer
{\it et al.},
Nature {\bf 440}, 315 (2006).


\bibitem[{\citenamefont{{Zaccanti} et~al.}(2009)\citenamefont{{Zaccanti},
  {Deissler}, {D'Errico}, {Fattori}, {Jona-Lasinio}, {M{\"u}ller}, {Roati},
  {Inguscio}, and {Modugno}}}]{Zaccanti:2009}
\bibinfo{author}{\bibfnamefont{M.}~\bibnamefont{{Zaccanti}}},
  \bibinfo{author}{\bibfnamefont{B.}~\bibnamefont{{Deissler}}},
  \bibinfo{author}{\bibfnamefont{C.}~\bibnamefont{{D'Errico}}},
  \bibinfo{author}{\bibfnamefont{M.}~\bibnamefont{{Fattori}}},
  \bibinfo{author}{\bibfnamefont{M.}~\bibnamefont{{Jona-Lasinio}}},
  \bibinfo{author}{\bibfnamefont{S.}~\bibnamefont{{M{\"u}ller}}},
  \bibinfo{author}{\bibfnamefont{G.}~\bibnamefont{{Roati}}},
  \bibinfo{author}{\bibfnamefont{M.}~\bibnamefont{{Inguscio}}},
  \bibnamefont{and}
  \bibinfo{author}{\bibfnamefont{G.}~\bibnamefont{{Modugno}}},
  \bibinfo{journal}{Nature Physics} \textbf{\bibinfo{volume}{5}},
  \bibinfo{pages}{586} (\bibinfo{year}{2009}).



  

\bibitem{Hammer:2010kp}
  H.-W.~Hammer, and L.~Platter,
  Ann.\ Rev.\ Nucl.\ Part.\ Sci.\  {\bf 60}, 207 (2010).


\bibitem{Canham:2008jd} 
  D.~L.~Canham, and H.~-W.~Hammer,
  Eur.\ Phys.\ J.\ A {\bf 37}, 367 (2008).
  
\bibitem{Canham:2009xg} 
  D.~L.~Canham, and H.~-W.~Hammer,
  Nucl.\ Phys.\ A {\bf 836}, 275 (2010).


\bibitem{Yamashita:2007ej} 
  M.~T.~Yamashita, T.~Frederico, and L.~Tomio,
  Phys.\ Lett.\ B {\bf 660}, 339 (2008).

\bibitem{Frederico:2012xh} 
  T.~Frederico, A.~Delfino, L.~Tomio and M.~T.~Yamashita,
  Prog.\ Part.\ Nucl.\ Phys.\  {\bf 67}, 939 (2012).
  



\bibitem{Rupakprl}
G.~Rupak and R.~Higa, Phys.\ Rev.\ Lett {\bf 106}, 222501 (2011)
%
\bibitem{Fernando:2011ts} 
  L.~Fernando, R.~Higa and G.~Rupak,
  Eur.\ Phys.\ J.\ A {\bf 48}, 24 (2012).
  
\bibitem{Hammer:2011ye} 
  H.~-W.~Hammer and D.~R.~Phillips,
  Nucl.\ Phys.\ A {\bf 865}, 17 (2011).
%
\bibitem{Rupak:2012}
  G.~Rupak, L.~Fernando and A.~Vaghani,
  Phys.\ Rev.\ C {\bf 86}, 044608 (2012).
 
\bibitem{Acharya:2013}
  B.~Acharya and D.~R.~Phillips,
  Nucl.\ Phys.\ A {\bf 913}, 103 (2013).

  
  \bibitem{Hagen:2013B}
    P.~Hagen, H.~-W.~Hammer and L.~Platter,
  Eur.\ Phys.\ J.\ A {\bf 49}, 118 (2013).
  
  
\bibitem{Zhang:2013kja} 
  X.~Zhang, K.~M.~Nollett and D.~R.~Phillips,
  Phys.\ Rev.\ C {\bf 89}, 024613 (2014).
 
  
\bibitem{Bertulani:2002sz} 
  C.~A.~Bertulani, H.~W.~Hammer, and U.~Van Kolck,
  Nucl.\ Phys.\ A {\bf 712}, 37 (2002).


\bibitem{Bedaque:2003wa} 
  P.~F.~Bedaque, H.~W.~Hammer, and U.~van Kolck,
  Phys.\ Lett.\ B {\bf 569}, 159 (2003).



\bibitem{Pascalutsa:2002pi} 
  V.~Pascalutsa and D.~R.~Phillips,
  Phys.\ Rev.\ C {\bf 67}, 055202 (2003).
  
\bibitem{Arndt:1973}
R.~A.~Arndt, D.~D.~Long, and L.~D. Roper,
Nucl.\ Phys. A {\bf 209}, 429 (1973). 

\bibitem{GHale}
G.~Hale, {\it private communication}.

\bibitem{Rotureau:2012yu} 
  J.~Rotureau and U.~van Kolck,
  Few Body Syst.\  {\bf 54}, 725 (2013).

 
\bibitem{Braaten:2011vf} 
  E.~Braaten, P.~Hagen, H.~-W.~Hammer and L.~Platter,
  Phys.\ Rev.\ A {\bf 86}, 012711 (2012).


\bibitem{Nishida:2011np} 
  Y.~Nishida,
  Phys.\ Rev.\ A {\bf 86}, 012710 (2012).

\bibitem{Jona-Lasinio:2008}
M.~Jona-Lasinio, L.~Pricoupenko, and Y.~Castin, Phys.\ Rev.\ A {\bf 77},  043611 (2008).


\bibitem[{\citenamefont{Birse et~al.}(1999)\citenamefont{Birse, McGovern, and
  Richardson}}]{Birse:1998dk}
\bibinfo{author}{\bibfnamefont{M.~C.} \bibnamefont{Birse}},
  \bibinfo{author}{\bibfnamefont{J.~A.} \bibnamefont{McGovern}},
  \bibnamefont{and} \bibinfo{author}{\bibfnamefont{K.~G.}
  \bibnamefont{Richardson}}, \bibinfo{journal}{Phys. Lett.}
  \textbf{\bibinfo{volume}{B464}}, \bibinfo{pages}{169} (\bibinfo{year}{1999}).

  
\bibitem[{\citenamefont{Kaplan et~al.}(1998)\citenamefont{Kaplan, Savage, and
  Wise}}]{Kaplan:1998tg}
\bibinfo{author}{\bibfnamefont{D.~B.} \bibnamefont{Kaplan}},
  \bibinfo{author}{\bibfnamefont{M.~J.} \bibnamefont{Savage}},
  \bibnamefont{and} \bibinfo{author}{\bibfnamefont{M.~B.} \bibnamefont{Wise}},
  \bibinfo{journal}{Phys. Lett.} \textbf{\bibinfo{volume}{B424}},
  \bibinfo{pages}{390} (\bibinfo{year}{1998}).


\bibitem[{\citenamefont{van Kolck}(1999{\natexlab{a}})}]{vanKolck:1998bw}
\bibinfo{author}{\bibfnamefont{U.}~\bibnamefont{van Kolck}},
  \bibinfo{journal}{Nucl. Phys.} \textbf{\bibinfo{volume}{A645}},
  \bibinfo{pages}{273} (\bibinfo{year}{1999}{\natexlab{a}}).
  

\bibitem{Beane:2000fx}
S.~R. Beane, P.~F. Bedaque, W.~C. Haxton, D.~R. Phillips, and
  M.~J. Savage,
 in {\it At the Frontier of Particle Physics: Handbook of QCD},
 edited by M. Shifman (World Scientific, Singapore, 2001), Vol. 1, p. 133.


\bibitem{Bedaque:2002mn}
P.~F. Bedaque, and U. van Kolck,
\newblock {Ann. Rev. Nucl. Part. Sci.}, {\bf 52}, 339 (2002).


\bibitem{GonzalezTrotter:2006wz} 
  D.~E.~Gonzalez Trotter
  {\it et al.},
  Phys.\ Rev.\ C {\bf 73}, 034001 (2006).
  
  
\bibitem{Slaus:1989}
I.~\v{S}laus, Y.~Akaishi, H.~Tanaka, 
Phys.\ Rept.\ {\bf 173}, 257 (1989).



\bibitem{Glockle:1983}
W.~Gl\"{o}ckle,
{\it The Quantum-Mechanical Few-Body Problem},
Springer-Verlag Berlin Heidelberg (1983).


\bibitem{Afnan:1977}
I.~R.~Afnan and A.~W.~Thomas, 
{\it Modern Three-Hadron Physics},
edited by A. W. Thomas (Springer-Verlag, Berlin, 1977), p. 1.

  

  
\bibitem{Faddeev:1960su}
  L.~D.~Faddeev,
  Sov.\ Phys.\ JETP {\bf 12}, 1014 (1961)
  [Zh.\ Eksp.\ Teor.\ Fiz.\  {\bf 39}, 1459 (1961)].

\bibitem{Arfken}
Arfken, G. ``Neumann Series, Separable (Degenerate) Kernels", Chapter 16.3 in {\it Mathematical Methods for Physicists}, 3rd ed., Academic Press, Orlando, FL (1985). 


\bibitem{Ji:2012he}
C.~Ji, C.~Elster, D.~Phillips,
{\it ``Universal correlations in the $^6$He ground state''}
in progress (2014).
  

\bibitem{Hammer:2001gh}
  H.~W.~Hammer, and T.~Mehen,
  Phys.\ Lett.\  B {\bf 516}, 353 (2001).


\bibitem{Ji:2011qg} 
  C.~Ji, D.~R.~Phillips, and L.~Platter,
  Annals Phys.\  {\bf 327}, 1803 (2012).
  

\bibitem{Ji:2012nj}
  C.~Ji and D.~R.~Phillips,
  Few Body Syst.\  {\bf 54}, 2317 (2013).
  
\bibitem{Vanasse:2013sda} 
  J.~Vanasse,
  Phys.\ Rev.\ C {\bf 88}, 044001 (2013).


\bibitem{Khaldi:2010jg} 
  K.~Khaldi, C.~.Elster, and W.~Gl\"{o}ckle,
  Phys.\ Rev.\ C {\bf 82}, 054002 (2010).
  
  \bibitem{Pieper:2004qw} 
  S.~C.~Pieper, R.~B.~Wiringa and J.~Carlson,
  Phys.\ Rev.\ C {\bf 70}, 054325 (2004).
  
\bibitem{Romero-Redondo:2013wma} 
  C.~Romero-Redondo, P.~Navr\'{a}til, S.~Quaglioni and G.~Hupin,
  Few-Body Syst. {\bf 55}, 927 (2014).

\bibitem{Tanihata:2008vw} 
  I.~Tanihata
{\it et al.},
  Phys.\ Rev.\ Lett.\  {\bf 100}, 192502 (2008).


\bibitem{Edmonds:1957}
A.~R.~Edmonds, 
{\it Angular Momentum in Quantum Mechanics}
(Princeton U.P., Princeton, N.J., 1957).

 
\end{thebibliography}
\end{document}